# Real-Time Denoising of Volumetric Path Tracing for Direct Volume Rendering

Jose A. Iglesias-Guitian,Prajita Mane and Bochang Moon *Member, IEEE*

**Abstract**—Direct Volume Rendering (DVR) using Volumetric Path Tracing (VPT) is a scientific visualization technique that simulates light transport with objects' matter using physically-based lighting models. Monte Carlo (MC) path tracing is often used with surface models, yet its application for volumetric models is difficult due to the complexity of integrating MC light-paths in volumetric media with none or smooth material boundaries. Moreover, auxiliary geometry-buffers (G-buffers) produced for volumes are typically very noisy, failing to guide image denoisers relying on that information to preserve image details. This makes existing real-time denoisers, which take noise-free G-buffers as their input, less effective when denoising VPT images. We propose the necessary modifications to an image-based denoiser previously used when rendering surface models, and demonstrate effective denoising of VPT images. In particular, our denoising exploits temporal coherence between frames, without relying on noise-free G-buffers, which has been a common assumption of existing denoisers for surface-models. Our technique preserves high-frequency details through a weighted recursive least squares that handles heterogeneous noise for volumetric models. We show for various real data sets that our method improves the visual fidelity and temporal stability of VPT during classic DVR operations such as camera movements, modifications of the light sources, and editions to the volume transfer function.

**Index Terms**—Volume rendering, global illumination, path-tracing, participating media, image-space filtering, real-time denoising.

## 1 Introduction

Recent studies evidenced perceptual benefits of applying more advanced illumination models for 3-D scientific visualizations [9], [12], [38]. Consequently, in the past years, interactive volume rendering techniques started supporting more advanced illumination effects [16], [41]. Direct Volume Rendering (DVR) using Volumetric Path Tracing (VPT) represents a new trend of volume rendering algorithms that use more advanced physically-based lighting models to produce photo-realistic scientific visualizations [8], [10], [49], [51]. This trend has also been popularized in medical imaging under the term Cinematic Rendering.

Global illumination models used in DVR are inspired by the radiative transfer equation [6], the fundamental equation governing light transport in participating media. Kajiya and Von Herzen [25] presented an approximate solution of this equation for its use in computer graphics. Monte Carlo (MC) path tracing, which is often used to solve this equation in an unbiased manner, has a unified theoretical framework that guarantees convergence to the exact solution. VPT computes DVR images by progressively averaging large numbers of radiance samples evaluated from randomly chosen light paths.

The major disadvantage of this algorithm is that generating high-quality DVR images requires large rendering times or tremendously expensive hardware equipment to achieve near-interactive framerates. Otherwise, the rendered images exhibit severe noise caused by MC integration of the samples. For example, Shih et al. [54] presented a parallelized, data-distributed and GPU-accelerated algorithm for volume rendering with advanced lighting. In particular, their method featured soft shadows and rendering on a cluster using up to 128 GPUs. Progressive MC volume rendering approaches, e.g., Exposure Render [33] or progressive light volumes [39], refine pixel colors using MC path tracing and tend to produce nearly noise-free images only after a few seconds. Nevertheless, these techniques still generate disturbing flickering noise while manipulating the camera or the transfer function, as real-time user interactions force the rendering to integrate only a reduced number of samples.

As a result, reducing noise and improving the temporal stability of DVR image sequences remains an open research problem [13]. In computer graphics, denoising for VPT has been explored mainly for offline production [4]. Instead, near-interactive or real-time denoising methods have mostly focused on scenes with surface models [5] and they have not been explored yet in the context of real-time DVR with heterogeneous participating media.

In this paper, we introduce new real-time denoising for DVR image sequences rendered using VPT. Our approach achieves real-time performance on commodity GPUs while reducing distracting MC noise and temporal flicker. Our high-level idea is to use image-space denoising, widely used when rendering surface models, with the necessary modifications that enable the denoiser to work effectively with VPT images.

The main contributions of this work are summarized as:

- We successfully introduce image-based denoising of VPT for real-time DVR. We propose a post-filtering technique that effectively reduces the noise level of VPT images, given only a few samples per pixel (spp).
- We extend a temporal denoiser based on recursive least squares (RLS) into a weighted RLS (wRLS) so that we can better handle

- *J.A. Iglesias-Guitian is with the Computer Vision Center, Universitat Autònoma de Barcelona, Bellaterra, 08193, Spain; and the Universidade da Coruña, CITIC - Centre for ICT Research, 15071, A Coruña, Spain. E-mail: See http://www.j4lley.com*
- *P. Mane and B. Moon (corresponding author) are with Gwangju Institute of Science and Technology, Gwangju, 61005, Republic of Korea. Email: prajitamane,bmoon@gist.ac.kr*

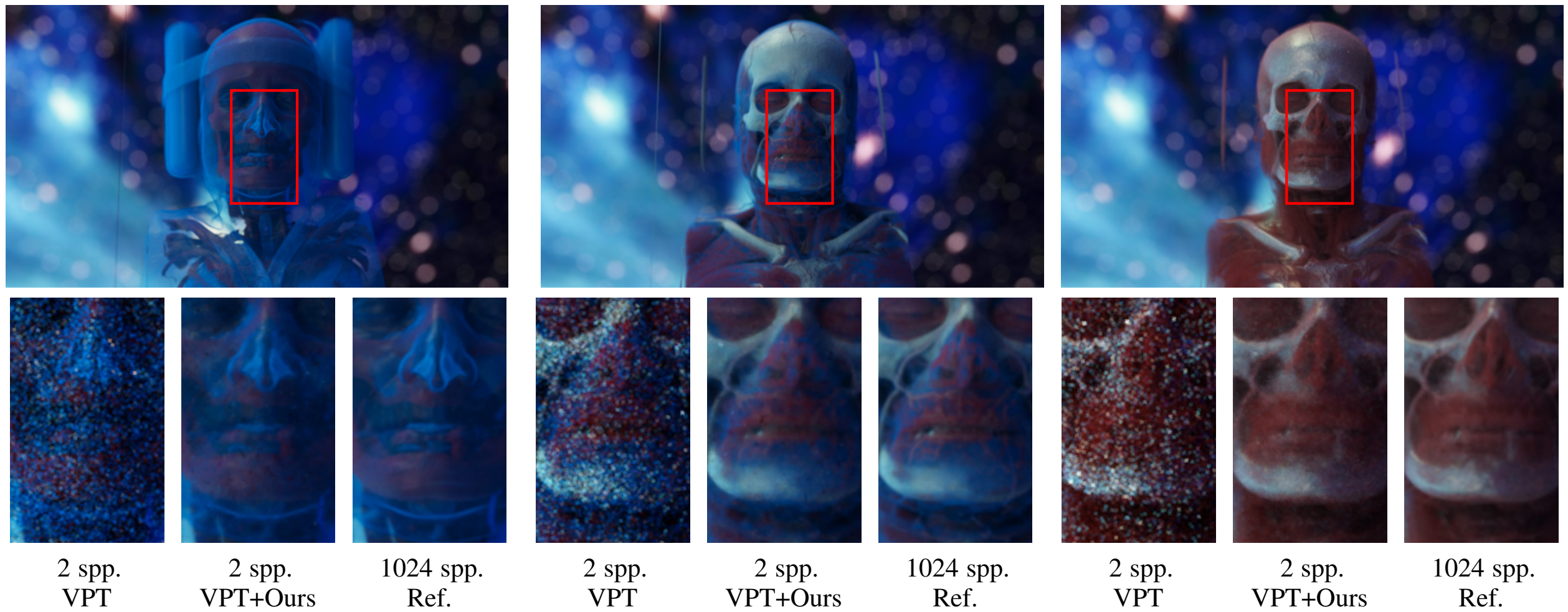

Figure 1: VPT results generated using the same source volume but during the interactive manipulation of different DVR transfer functions. Multiple scattering bounces per ray are simulated. Our real-time denoising improves VPT images (MC-DVR with only 2 spp) while reducing its noise effectively. Offline VPT with 1024 spp, taking minutes to produce a single image, is shown as reference.

the heterogeneous noise of VPT by controlling the denoising weight assigned to each pixel color.

- We demonstrate that our real-time denoiser improves the numerical accuracy of rendered images while reducing the temporal flicker when a user manipulates parameters of cameras, light sources, and volume transfer functions (e.g., Fig. 1). In particular, our method relies on neither G-buffers nor any pre-training.

## 2 Related Work

The underlying physical assumptions of the various optical models used for light transport simulation in participating media were reviewed in Max et al. [43], [44]. We also refer to Jönsson et al. [23] for a comprehensive survey on interactive volume rendering. In this section, we mainly discuss the existing approaches for global illumination on volume rendering with a particular emphasis on interactive and progressive techniques.

### 2.1 Monte Carlo path tracing

The rendering equation [25] for participating media can be solved using path tracing algorithms [37]. The recent survey by Novák et al. [46] reviewed the latest advances in MC path tracing methods to solve the light transport in participating media. Direct Volume Rendering (DVR) techniques in scientific visualization have typically employed ray marching algorithms. For instance, Rezk-Salama [52] proposed an interactive GPU-based MC ray-casting approach for physically-based volume rendering that used ray-marching. Ray marching is simple but has several drawbacks. For example, it tends to be expensive for high-resolution volumes, and the rendered images can exhibit an unpredictable bias since high-frequency details can be missed [13], [47].

**Improved sampling strategies.** An alternative to the ray marching is delta tracking, which is an importance sampling that determines free paths according to the probability density function (PDF) corresponding to the optical depth in the participating medium. Woodcock tracking [7] is a widely adopted unbiased solution that adjusts sampling distances to be small enough to sample dense regions in the volume appropriately. This algorithm has been revisited in offline rendering for adaptive sampling on large sparse inhomogeneous media [59] and further optimized for film production [35]. Free path sampling with probabilities not necessarily proportional to the volume transmittance has been realized using weighted delta tracking approaches [36], [47], [56]. All these techniques help reducing noise in the estimated light paths for participating media. Our real-time denoising can be complementary to the underlying sampling techniques.

**Progressive MC path tracing.** Kroes et al. [33] and Liu et al. [39] demonstrated that progressive VPT using GPUs could achieve interactive frame rates for unbiased volume rendering. Unfortunately, while progressive VPT can converge to noise-free images, it comes with the penalty of producing very noisy results for interactive rendering scenarios or requiring expensive cloud-based or distributed rendering systems [49], [54]. We alleviate this problem of noise in progressive VPT by applying our real-time denoising as a post-processing.

**Image denoising for MC path tracing.** Image-space reconstruction has been widely accepted as a viable alternative to reduce MC path tracing noise in surface models. A comprehensive survey on the topic was conducted by Zwicker et al. [62]. Recently, Schied et al. [53] presented real-time spatiotemporal denoising that accumulates pixel colors across frames and controls its smoothing level using the variances of the colors. Mara et al. [42] designed a real-time denoiser tailored to reducing noise in matte and glossy surfaces. These techniques are specialized for filtering noisy global illumination for surface models. The denoising for volumes and participating media has often been related to production offline rendering scenarios [4]. Deep learning denoisers have recently gained much popularity [2], [14]. For example, Chaitanya et al. [5] proposed an interactive denoiser with deep learning for MC path tracing. However, these denoisers for surface models heavily rely on noise-free G-buffers, often unavailable for volumes, to produce high-quality denoised images.

In the context of DVR, Kroes et al. [33] applied a general noise reduction filter [27] as part of their GPU implementation, but this general filtering did not succeed in effectively removing MC variance and temporal flicker. Applying specialized denoising for surface models (e.g., RLS adaptive denoising [45]) to VPT can be ineffective due to the noise in the G-buffer information used to reconstruct image details. Consequently, we design our denoising to reduce variance noise and temporal flicker for interactive DVR using VPT, without relying on the problematic G-buffers. This allows our technique to produce temporally stable results without any pre-training for different given types of user interactions. It differentiates our method from existing denoisers relying on G-buffers and also from deep-learning-based approaches using expensive pre-training stages.

### 2.2 Irradiance caching: precomputed radiance transfer

Irradiance caching takes advantage of smoothly varying indirect illumination and precomputes radiance transfer inside the volume. In an early work, Kajiya and Von Herzen [25] proposed a two-pass approach that simulates global illumination effects for heterogeneous volume datasets. On the first pass, the radiance is estimated in each voxel and consecutively integrated along view rays in the second pass. The first pass, however, is time-consuming and thus not applicable to interactive visualization. Alternatively, the estimated irradiance can be computed only at a sparse set of cached points in the volume. For example, Kriványek et al. [30], [31], [32] proposed storing and interpolating direction-dependent radiance using spherical harmonics (SH). Later, Jarosz et al. [18] extended this approach for participating media. Kronander et al. [34] obtained real-time performance for DVR by encoding local and global volumetric visibility with SHs on a multi-resolution grid. More recently, Khlebnikov et al. [28] proposed parallel irradiance caching with MC path tracing for interactive volume rendering. While irradiance caching stores and updates precomputed radiance, our denoising approach does not require any pre-processing.

### 2.3 Volumetric photon mapping approaches

Volumetric photon mapping [21] and progressive extensions [15] amortized expensive calculations to solve the volume rendering integral through caching light-transport. Jarosz et al. introduced a variation of Woodcock tracking in progressive photon beams [19], an extension of photon beams [20] for volumetric photon mapping. Jönsson et al. [22] realized interactive DVR through photon mapping by recomputing only the photons that have changed. However, their photon gathering stage was computationally expensive, leading to low frame rates when the camera moves. Zhang et al. [61] proposed a precomputed volume radiance transfer using precomputed photon maps encoded using basis functions. While this method allows real-time radiance reconstruction, the photon map should be regenerated every time the transfer function changes. To accelerate the photon map generation, Jönsson et al. [24] identified photons invariant to changes of visual parameters (e.g., changes in the transfer function), enabling a further reduction of the overhead associated with recomputing photon maps. Unlike MC path tracing, photon mapping typically introduces some bias, often visible as low-frequency noise.

### 2.4 Many-light methods

Many-light methods, as photon mapping, are both bidirectional MC techniques. While photon mapping relies on density estimation and requires a large number of photons to be traced, many-light methods require orders of magnitude less light paths, and thus rendering can be very efficient. Engelhardt et al. [11] described a particle tracing algorithm to create a set of Virtual Point Lights (VPLs) within participating media and derived a GPU-friendly bias compensation scheme for high-quality rendering. Weber et al. [57] applied a many-light approach using VPLs to interactive volume rendering. In particular, this technique was tailored for interactive editing of volume transfer functions, providing immediate updates and redistribution of the contributions from VPLs. However, interactive visualization restricts the number of VPLs, and transfer-function edits are limited to smooth transitions as they require further redistribution and recomputation of VPLs. Temporal coherence was improved by progressively updating the positions of VPLs and refreshing only their incremental contributions. However, significant changes (e.g., switching to a completely different transfer function) could cause visible flickering. Our approach is able to address these changes in the transfer function without noticeable flicker.

### 2.5 Diffusion approximations

For the rendering of multiple scattering global illumination effects in participating media, methods based on the diffusion approximation [55] are an efficient alternative to MC path tracing. Körner et al. [29] proposed Flux-Limited Diffusion (FLD), a technique improving over Classical Diffusion Approximations (CDA) for heterogeneous media. While CDA methods suffer from non-physical radiative fluxes in transparent regions, FLD produces more accurate results than CDA when compared to the path traced ground truth. Although the proposed FLD solver can converge faster than MC path tracing or photon mapping, no progressive or interactive extensions have yet been proposed.

## 3 THE VOLUME RENDERING INTEGRAL

This section explains the fundamentals of progressive VPT to obtain MC solutions of the Volume Rendering Integral (VRI).

TABLE 1: Notations for light-matter interactions.

| Symbol | Description |
|---|---|
| $\omega$ | direction vector of light propagation |
| $\mathbf{x}, \mathbf{y}, \mathbf{z}$ | boldface represents 3-D sampling points |
| $x, y, z$ | italicized represents distances from 3-D points to the ray origin (e.g., if the ray origin is at $\mathbf{x}$, then $x = 0$) |
| $\rho(\cdot)$ | probability density function (PDF) |
| $P(\mathbf{z})$ | probability of sampling at point $\mathbf{z}$ |
| $\mu_a(\mathbf{y}), \mu_s(\mathbf{y})$ | absorption and scattering probabilities at point $\mathbf{y}$ |
| $\mu_t(\mathbf{y})$ | extinction coefficient at point $\mathbf{y}$ |
| $T(\mathbf{x}, \mathbf{y})$ | transmittance or attenuation of light between two points |
| $L_e(\mathbf{x}, \omega)$ | emission energy at point $\mathbf{x}$ in the direction $\omega$ |
| $L_s(\mathbf{x}, \omega)$ | in-scattering energy at point $\mathbf{x}$ coming from direction $\omega$ |
| $L(\mathbf{z}, \omega)$ | emitted or reflected energy at point $\mathbf{z}$ in the direction $\omega$ coming from a background surface |
| $\bar{\mathbf{x}}$ | a light transport path |
| $f(\bar{\mathbf{x}})$ | contribution of the differential flux carried by path $\bar{\mathbf{x}}$ |

Following similar mathematical notations used in Novák et al. [46] (see Table 1), we can write the general VRI as follows (more details are in our supplementary report):

$$L(\mathbf{x}, \omega) = \int_0^z T(\mathbf{x}, \mathbf{y}) \left[ \mu_a(\mathbf{y}) L_e(\mathbf{y}, \omega) + \mu_s(\mathbf{y}) L_s(\mathbf{y}, \omega) \right] \mathrm{d}y + \underbrace{T(\mathbf{x}, \mathbf{z}) L(\mathbf{z}, \omega)}_{background} \quad (1)$$

Iterative approximations of the VRI can be achieved through progressive VPT, computing only a small number of light transport trajectories (*light paths*) in a single frame (see Fig. 2). So, the $j$-th pixel color in the rendered image $I$ can be represented as the following integral:

$$I_j = \int_{\mathscr{P}} f_j(\bar{\mathbf{x}})\,\mathrm{d}\bar{\mathbf{x}}, \quad (2)$$

where $\mathscr{P}$ is the space of all possible light paths in the scene. Applying MC estimation to the integral in Eq. 1, we can obtain:

$$\langle L(\mathbf{x},\omega)\rangle = \frac{T(\mathbf{x},\mathbf{y})}{\rho(\mathbf{y})}\left[\mu_a(\mathbf{y})L_e(\mathbf{y},\omega) + \mu_s(\mathbf{y})L_s(\mathbf{y},\omega)\right] + \frac{T(\mathbf{x},\mathbf{z})}{P(\mathbf{z})}L(\mathbf{z},\omega). \quad (3)$$

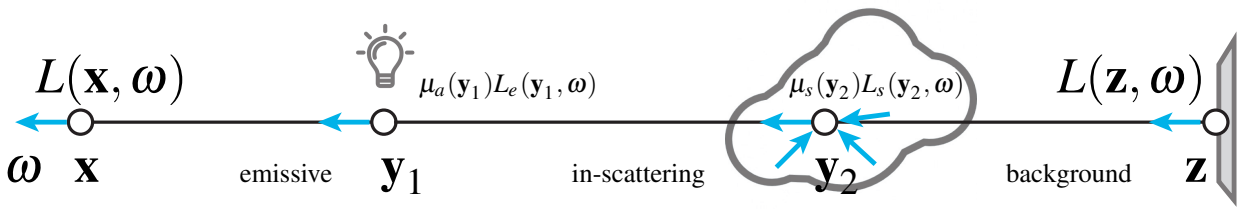

Figure 2: Illustration of an exemplar light path $\bar{\mathbf{x}}$, transporting $L(\mathbf{x}, \omega)$ radiance energy to its ray origin at $\mathbf{x}$ in the direction $\omega$.

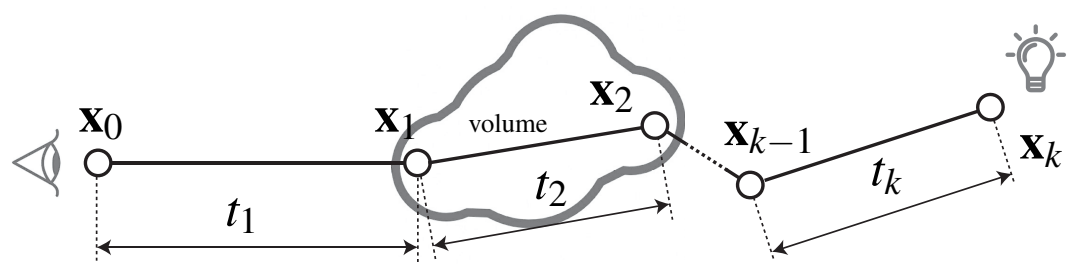

Figure 3: Building incremental light paths. One way to avoid building excessively long paths is to place the last vertex $\mathbf{x}_k$ inside a randomly selected light source after only a few real collisions.

The main advantage of this approach is that it only requires to evaluate one path segment per light path at a time.

**Incremental light transport paths.** A common approach for constructing a light transport path $\bar{\mathbf{x}} = (\mathbf{x}_0, \mathbf{x}_1, \ldots, \mathbf{x}_k) \in \mathscr{P}$ is to start from the camera at point $\mathbf{x}_0$ and extend the path incrementally segment by segment (see Fig. 3). To determine the location of the next vertex $\mathbf{x}_{i+1}$ of a light path, a ray direction $\omega_{i+1}$ is sampled with $\rho(\omega_{i+1})$, which depends on the medium phase function or the surface bidirectional scattering distribution function (BSDF) at $\mathbf{x}_i$. For this work, we considered only isotropic scattering where the phase function is $\frac{1}{4\pi}$ for all directions.

**Distance sampling.** The iterative evaluation of Eq. 3 requires to find a suitable discrete position $\mathbf{y}$, and evaluate its illumination contribution. Delta tracking for free-path sampling utilizes the concept of null collisions with a fictitious matter to achieve correct sampling on heterogeneous volumes. It finds the candidate position $\mathbf{y}$ by recursively sampling tentative collisions until one classified as real is found. Real collisions are accepted with $\rho(\mathbf{y})$ proportional to the extinction coefficient $\mu_t(\mathbf{y})$. While null collisions should not affect light transport, they require expensive memory accesses to evaluate extinction coefficients in the volume. One general strategy to generate a tight extinction bound is to sample free-paths using uniform random numbers $\xi \in [0,1)$. A candidate lower distance threshold can be therefore defined as:

$$t(\xi) = -\frac{ln(1-\xi)}{\mu_t^{max}}, \quad (4)$$

where $\mu_t^{max}$ is the maximum extinction coefficient in the volume. Once delta tracking classifies a tentative collision as real and it absorbs more energy than the bound in Eq. 4, then $\mathbf{y}$ is considered as the first actual collision. One advantage of using delta tracking for DVR is that it is aware of transfer function alterations since each collision is tested after transfer function values have been applied to the volume. Improved distance sampling strategies for interactive volume rendering would represent an orthogonal research direction to this work.

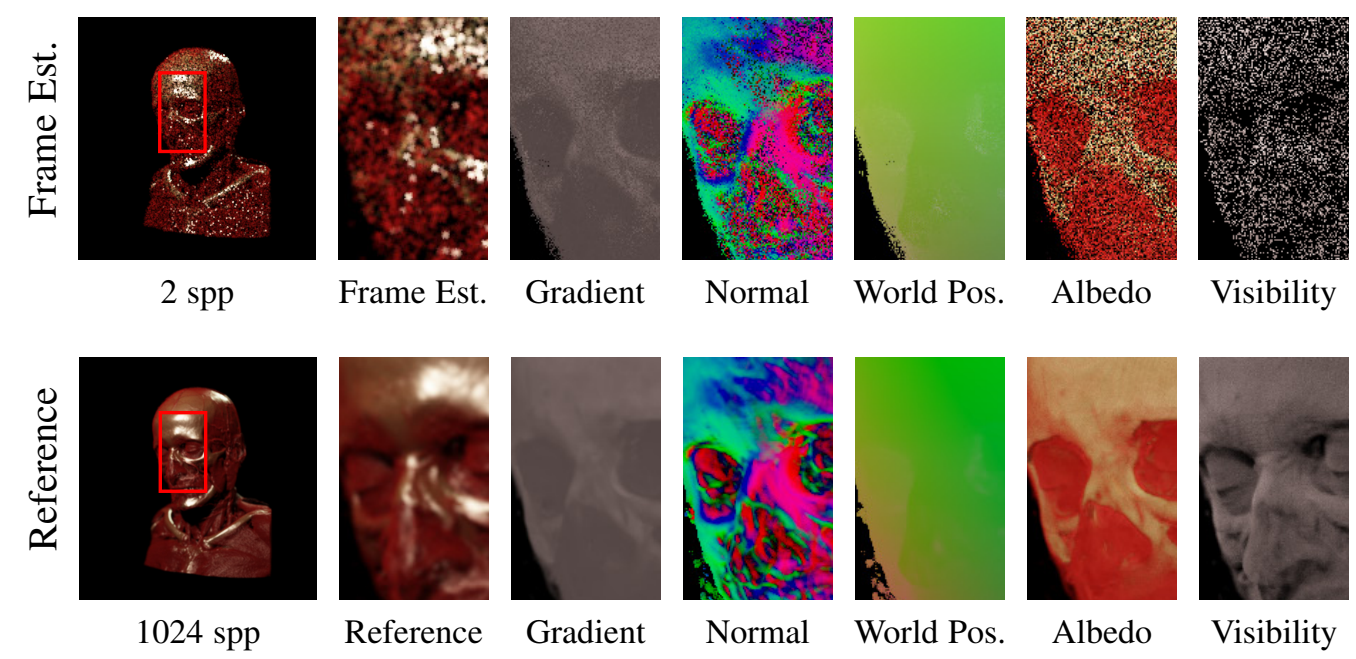

Figure 4: Visualization of the auxiliary buffers commonly used in image-space denoising. It can be noticed that the buffers computed with 2 spp are very noisy. For illustrative purposes, we show reference features computed using 1024 spp.

## 4 Spatio-Temporal Denoising Framework

Using VPT is an attractive choice to support physically-based global illumination on DVR frameworks, as the VPT is a simple and general algorithm that covers a wide variety of lighting effects. However, VPT tends to produce noisy approximations of the VRI due to its stochastic nature, especially given a limited number of samples under the real-time constraints of DVR. In particular, VPT noise degenerates into temporal flicker when a user interacts with the parameters of a DVR scene (e.g., lighting or transfer functions), since VPT generates a new image from scratch, leading to lower image quality. Our primary goal is to handle such noise in VPT results through a real-time denoising framework that takes advantage of spatial and temporal coherence among pixel colors to obtain numerically and visually improved high-quality interactive DVR results.

**Challenges for DVR image-based denoising.** Recent image denoising methods [62] often utilize auxiliary features, also known as G-buffer features (e.g., depth, normal, and albedo), to preserve high-frequency information in rendered images. The features can be much less noisy than radiance values for surface models, but those can be extremely noisy, as shown in Fig. 4, when rendering volumetric models with a small sample count (e.g., 1 or 2 spp) under real-time constraints. Our denoising technique shares similarities with previous image denoisers (e.g., [17], [45]) in the sense that both reduce MC variance by blending pixel colors as a weighted sum. However, the key difference is that our method exploits a stable feature formed by accumulating pixel colors over time, instead of relying on the G-buffers, unlike the recent denoisers specialized for surface models.

**Real-time denoising framework.** Our real-time framework (Fig. 5) is built upon progressive VPT for DVR. At each frame, MC-DVR estimates the amount of radiance arrived at each pixel based on VPT and the randomly generated light paths. The resulting distribution of radiance is generally a noisy estimate given the low sample count imposed by real-time rendering constraints. Each estimated pixel color is then filtered using a pixel reconstruction filter to obtain the current frame estimate. Our DVR denoising technique takes the current frame estimate and updates a temporal denoising feature that exploits temporal coherence and guides our denoiser. Finally, our denoising technique will obtain the final image using per-pixel linear model predictions. Our framework supports both spatial and temporal filtering using those predictions.

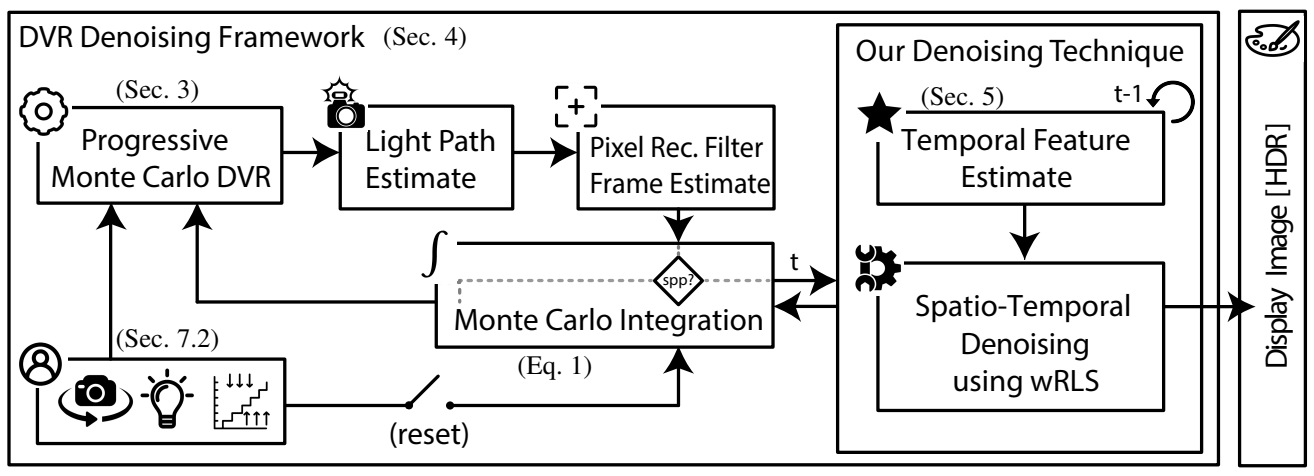

Figure 5: Real-time DVR denoising framework. Our real-time denoising is integrated into the DVR framework as postprocessing of VPT. We take noisy frame estimates as input and reduce the noise by utilizing spatial and temporal coherence among pixel colors through per-pixel linear models, which utilize temporally generated per-pixel features.

# 5 Image-space DVR denoising technique

In this section, we propose our real-time image-space denoiser for DVR images rendered using VPT. We first describe an approximation of the radiance arriving at each pixel of a DVR image using linear models (Sec. 5.1), then present RLS that estimates the coefficients of linear models in an online manner (Sec. 5.2). Lastly, we propose a new spatio-temporal denoiser using a weighted RLS, which reduces temporal flickering when rendering DVR images using VPT with low sample counts (Sec. 5.3).

TABLE 2: Notations for linear model predictions

| Symbol | Description |
|---|---|
| $I$ | ground-truth values of an image |
| $\tilde{I}$ | MC estimated values for an image with a virtual sensor |
| $\hat{I}$ | predicted values of an image $I$ using linear models |
| $I_j(t)$ | $j$-th pixel color of an image $I$ at time $t$ |
| $\mathbf{p}_j(t)$ | predictor vector at time $t$ for pixel $j$ used for linear regression |
| $\mathbf{z}_j(t)$ | feature vector at time $t$ for pixel $j$ |
| $\beta_j(t)$ | linear model coefficients for pixel $j$ at time $t$ |
| $\upsilon_j$ | velocity vector $(\upsilon_x, \upsilon_y)$ at pixel $j$ used for reprojection |
| $\pi(\upsilon_j)$ | reprojection operation $\pi(\upsilon_j) = j + \upsilon_j$ |
| $\xi_j(t)$ | single-channel prediction error for pixel $j$ at time $t$ |
| $e_j(t)$ | real error for pixel $j$ at time $t$ considering color channels |
| $\hat{e}_j(t)$ | estimated error for pixel $j$ at time $t$ considering color channels |

**Notation.** Let us define useful terms and notations used throughout the rest of the paper (see Table 2). Given a virtual camera sensor containing $n$ pixels that ideally could capture a MC-DVR ground truth image $I$, for each $j$-th pixel, $j \in [1,n]$, the element $I_j$ receives the contributions of one or multiple light paths $f(\bar{\mathbf{x}})$. All estimated light paths using VPT and contributing to pixel $j$ are denoted as $f_j(\bar{\mathbf{x}})$, and they are filtered using a pixel reconstruction filter to obtain a discrete real measurement $\tilde{I}_j$, an estimate of the MC integral (Eq. 2). We use $\tilde{I}_j(t)$ and $\hat{I}(t)$ to refer, respectively, to the noisy MC estimate and denoised value for the $j$-th pixel color at time $t$. We denote our temporal feature guiding our linear model predictions as $\mathbf{z}$.

## 5.1 Radiance estimation using linear models

We model the ground truth color $I_j$ at $j$-th pixel using the following linear regression:

$$I_j = \mathbf{p}_j \beta_j^T + \xi_j, \tag{5}$$

where $\mathbf{p}_j$ and $\beta_j$ represents the input predictor vector and its coefficients with length $d$, respectively. $\xi_j$ represents the prediction error of the linear regression model $\mathbf{p}_j \beta_j^T$. Linear models are used to find linear dependencies between the input regressors $\mathbf{p}_j$ and the ground truth signal $I_j$. Usually $\mathbf{p}_j \equiv [1, \mathbf{z}_j]$, where the first element of $\mathbf{p}_j$ is the intercept term and $\mathbf{z}_j$ is a feature vector of the model. For brevity's sake, we shall treat the value $I_j$ as a scalar unless otherwise mentioned, since our denoising is applied to each color channel independently. Note that the linear model represents an approximation of the integral (Eq. 2) over all light paths $f(\bar{\mathbf{x}})$ for the $j$-th pixel:

$$\int_{\mathscr{P}} f_j(\bar{\mathbf{x}})\, \mathrm{d}\bar{\mathbf{x}} \approx \hat{I}_j = \mathbf{p}_j \beta_j^T. \tag{6}$$

In an interactive context, the radiance value (i.e., pixel color) $I_j$ can vary over time, and thus we linearly model its temporal change using the predictor $\mathbf{p}_j$. More specifically, the values from $I_j(t-\delta)$ at frame $t-\delta$ to $I_j(t)$ at frame $t$, can be estimated as:

$$\begin{pmatrix} \hat{I}_j(t) \\ \hat{I}_j(t-1) \\ \vdots \\ \hat{I}_j(t-\delta) \end{pmatrix} = \underbrace{\begin{pmatrix} \mathbf{p}_j(t) \\ \mathbf{p}_j(t-1) \\ \vdots \\ \mathbf{p}_j(t-\delta) \end{pmatrix}}_{\mathbf{X}_j} \underbrace{\begin{pmatrix} \beta_{j,0} \\ \beta_{j,1} \\ \vdots \\ \beta_{j,d-1} \end{pmatrix}}_{\beta_j^T}. \tag{7}$$

In the equation above, $\mathbf{X}_j$ is the design matrix that concatenates the predictor vectors $\mathbf{p}_j(t-\delta), \ldots, \mathbf{p}_j(t)$ over time. Next, we describe Recursive Least Squares (RLS), a method to estimate the model coefficients $\beta$. After that, we will propose our own adaptation of RLS by estimating alternative input predictor vectors, and how to handle heterogeneous VPT noise more robustly.

## 5.2 Linear model regression using RLS

The high-level approach of RLS [40] is to update the coefficients of the statistical models based on differences between model predictions and measured values in an online manner. Ideally, for a given pixel $j$ at frame $t$, we would compute the real error by using the values of the ground truth image $I_j$, as $e_j(t) = I_j(t) - \hat{I}_j(t)$. Because the ground truth of the MC integral is not available in practice, we need to estimate this error using the noisy MC estimate $\tilde{I}_j(t)$ as the following:

$$\hat{e}_j(t) = \tilde{I}_j(t) - \hat{I}_j(t) = \tilde{I}_j(t) - \mathbf{p}_j(t)\beta_j^T(t-1), \tag{8}$$

where $\mathbf{p}_j(t)$ is the predictor vector concatenating the auxiliary features $\mathbf{z}_j(t)$. For example, interactive denoising methods using RLS (e.g., [45]) typically exploits noise-free G-buffers as the auxiliary features. Given the residual $\hat{e}_j(t)$, the model coefficients $\beta_j$ are incrementally updated at time $t$ [40]:

$$\beta_j(t) = \beta_j(t-1) + \mathbf{q}_j(t)\hat{e}_j(t) \tag{9}$$

$$\mathbf{q}_j(t) = \frac{\mathbf{P}_j(t-1)\mathbf{p}_j^T(t)}{\lambda + \mathbf{p}_j(t)\mathbf{P}_j(t-1)\mathbf{p}_j^T(t)}, \tag{10}$$

where $\lambda$ is the forgetting factor, typically fixed to a value near one (e.g., $\lambda = 0.998$), and $\mathbf{P}_j(t-1)$ is a $d \times d$ matrix which contains the inverse covariance of the predictor vectors. The inverse covariance matrix is updated at frame $t$ using the matrix inversion lemma [40] and the predictor vector $\mathbf{p}_j(t)$:

$$\mathbf{P}_j(t) = \lambda^{-1}\left(\mathbf{P}_j(t-1) - \mathbf{q}_j(t)\mathbf{p}_j(t)\mathbf{P}_j(t-1)\right). \quad (11)$$

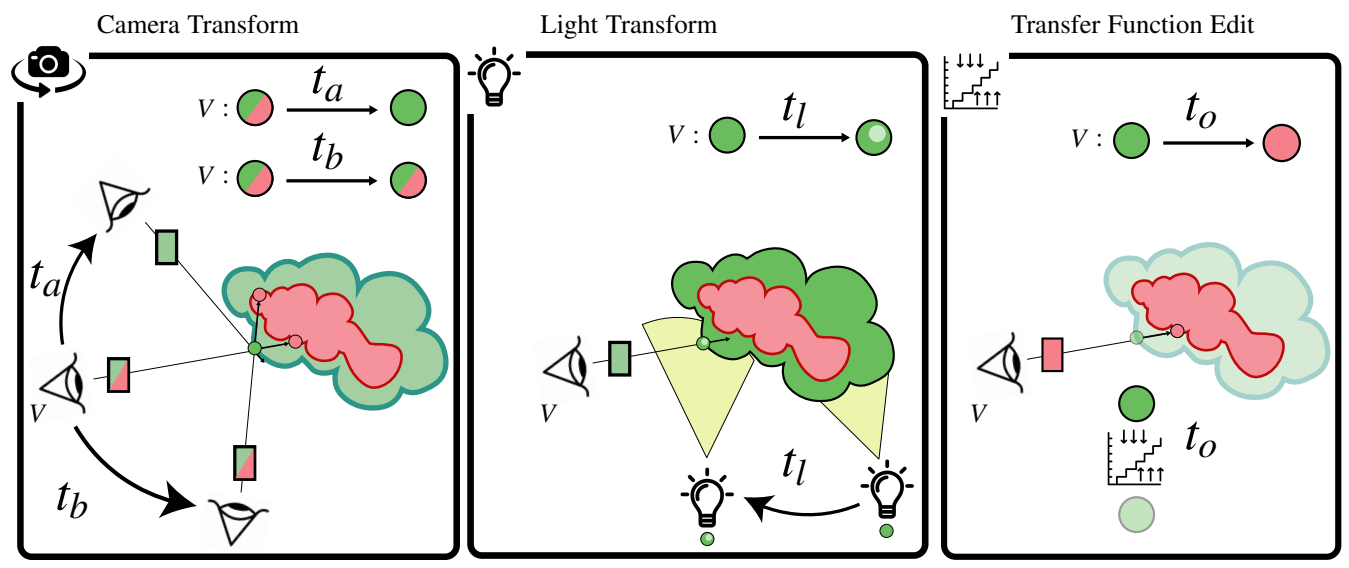

Figure 6: Reprojection on volumetric participating media. Linear model reprojection reacts to camera transformations (i.e. $t_a$ or $t_b$), light source modifications ($t_l$), and edits of the volume transfer function ($t_o$). This adaptative behavior of linear models avoids triggering additional MC integration reset operations.

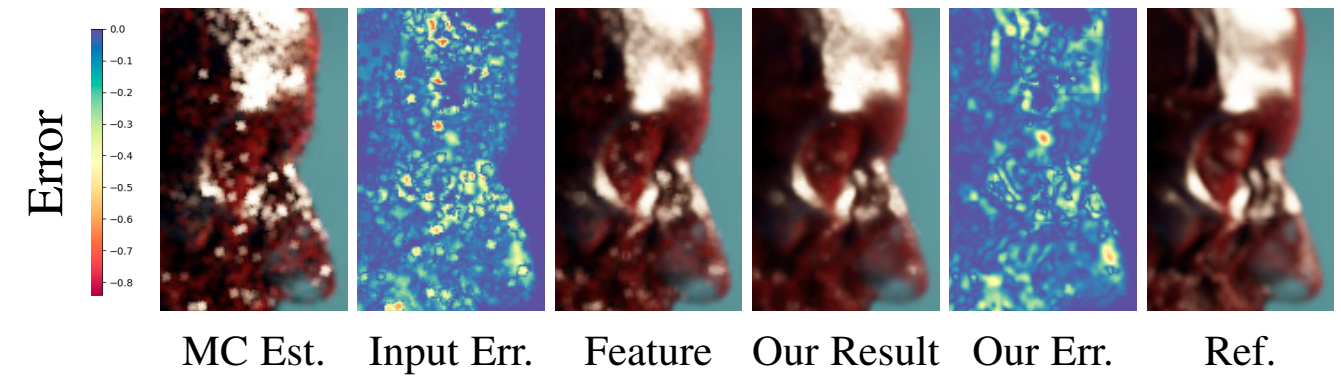

Figure 7: Visualization of the VPT input, our denoising feature, our denoised result and the reference. We also display the corresponding error maps for the input and our final denoised result. We display frame #100 of the MANIX camera animation.

### 5.3 Our proposed denoising for MC-DVR: wRLS

The problems of the RLS approaches when facing stochastic MC-DVR are still double. First, VPT does not generate noise-free G-buffers when rendering with low sample counts, and thus utilizing the buffers for the predictor vector is problematic. Second, the noise level of the input estimates $\tilde{I}_j(t)$ can vary significantly over time and can result in high variability of the linear model coefficients. These challenges caused by real-time stochastic VPT were not addressed in the previous RLS techniques [17], [45]. In this section, we propose a weighted RLS with a temporally stable feature to tackle the challenges.

**Temporal coherence of linear models for VPT.** Temporal reprojection is a well-known technique to exploit temporal coherence in consecutive frames, and it maintains an additional storage (e.g., history buffer) where pixel colors are accumulated over time.

While the conventional temporal reprojection is to reproject pixel colors, our method reprojects linear models to exploit the temporal coherence more robustly for real-time VPT. Reusing linear models has several advantages over reprojection schemes based on caching constant values per pixel (e.g., a history buffer). Linear regression can predict gradual changes in the camera, light sources and transfer functions, but also it reacts immediately to abrupt changes affecting shading (see Fig. 6). We estimate per-pixel velocities $\upsilon_j$ (i.e., optical flow) of a linear model using the view matrix and per-pixel world coordinates. Implementation details for computing the optical flow will be given in Sec. 6. Once a velocity $\upsilon_j$ for the $j$-th pixel is calculated, we can define a reprojection operation $\pi$ that obtains the corresponding pixel coordinates $q$ in the precedent frame as $q \leftarrow \pi(\upsilon_j)$. Even for cases where $\upsilon_j = 0$, like light or transfer functions changes, linear models can predict gradual changes.

**Temporal denoising feature.** We propose a temporal feature for our denoising, without relying on noisy G-buffers. Ideally, the feature should have a low variance in the temporal dimension and have a high correlation with the ground truth image. To this end, we adopted an exponentially weighted history buffer (e.g., [17]) as the feature $\mathbf{z}_j \in \mathbb{R}^3$ of our linear model at pixel $j$. Specifically, the feature, which is corresponding to the pixel color (e.g., RGB values) in the history buffer, is updated at frame $t$ given the MC estimates $\tilde{I}(t)$:

$$\mathbf{z}_j(t) = \alpha\Psi[\mathbf{z}_q(t-1), \tilde{I}_j(t)] + (1-\alpha)\tilde{I}_j(t), \quad (12)$$

where $\alpha$ is the weight that controls the balance between history and the current estimate. Note that the pixel index $q$ in the previous frame can be different from the current index $j$, and that $q$ is computed using the per-pixel velocity as $\pi(\upsilon_j)$. The reprojected history $\mathbf{z}_q(t-1)$ is rectified using $\psi[\cdot]$, the neighborhood clamping operator [26] that relies on the current-frame input estimate, $\tilde{I}_j(t)$, to reduce artifacts caused by stale history data. Even though we denoise color channels independently, our feature $\mathbf{z}_j(t)$ is constructed using the three channels simultaneously to avoid undesired color-shift effects. Fig. 7 shows an example feature image that has much reduced noise compared to the input image. Given the features constructed by accumulating colors over time, we introduce a variant of RLS (i.e., weighted RLS) that is able to handle temporally varying noise.

**Weighted recursive least squares (wRLS).** As a key technical contribution, we propose a weighted RLS (wRLS) that takes into account the heterogeneous noise in DVR images generated with small numbers of samples. At a high-level, the wRLS allocates high weights to samples with low variance and low weights to samples with high variance. In particular, a very low weight is assigned to outlier samples, which have an extremely high variance so that our linear models can produce temporally stable results. The straightforward way to compute the weight is to utilize sample variances of pixel colors, but this cannot be robustly achieved for our real-time scenarios where only a few samples are available. To tackle this challenge, we control the weight by exploiting our temporally stable feature. Explicitly, the weight $w_j(t)$ assigned to $\tilde{I}_j(t)$ is computed as:

$$w_j(t) = e^{-d_j^2(t)/h^2}, \quad d_j(t) = \frac{\|\tilde{I}_j(t) - \mathbf{z}_j(t)\|}{min(\|\tilde{I}_j(t)\|, \|\mathbf{z}_j(t)\|) + \varepsilon}, \quad (13)$$

where $\varepsilon$ is a very small number to avoid divisions by zero, and $h$ is a filtering bandwidth that controls a tradeoff between denoising bias and variance. For example, smaller $h$ values would provide a more temporally stable but higher bias. We found that $h = 0.75$ produces a good balance between the temporal stability and bias for our tests. The modified equation to update linear models using this weight is:

$$\mathbf{q}_j^w(t) = \frac{\mathbf{P}_j(t-1)\mathbf{p}_j^T(t)}{\frac{\lambda}{w_j(t)} + \mathbf{p}_j(t)\mathbf{P}_j(t-1)\mathbf{p}_j^T(t)}. \quad (14)$$

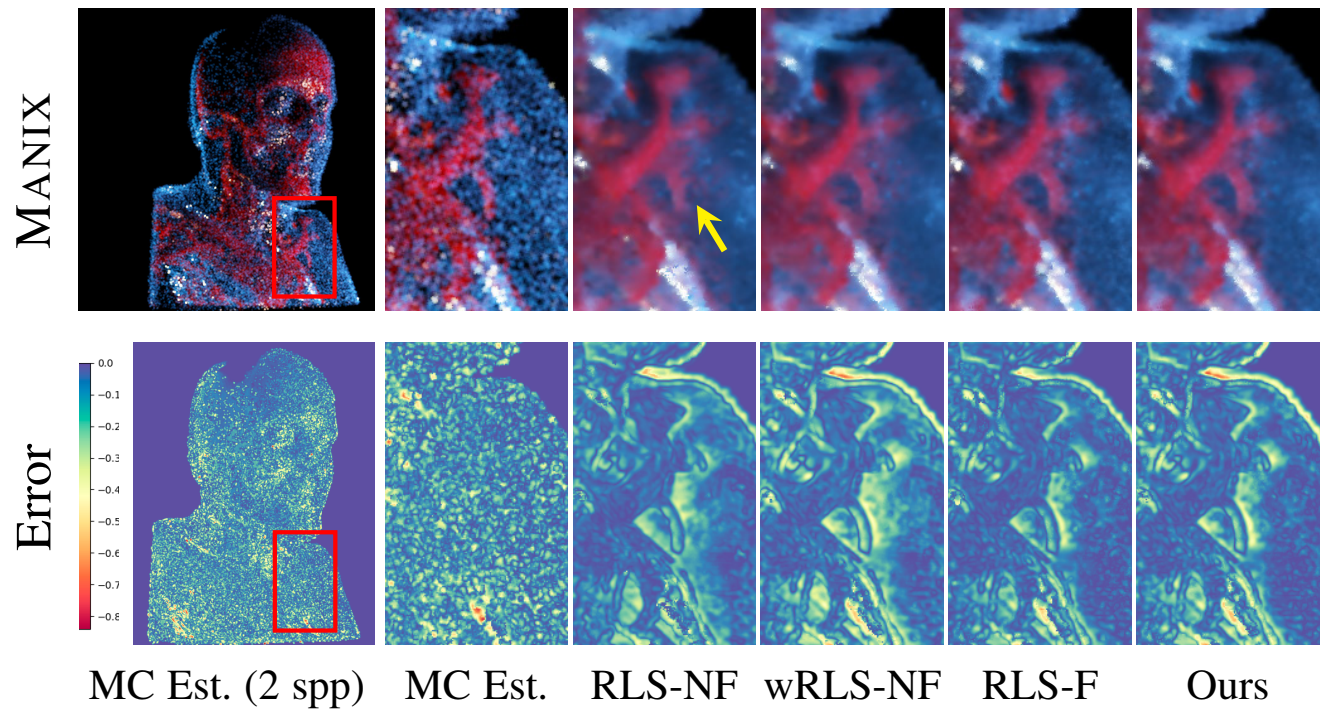

Figure 8: Ablative comparisons of our contributions for denoising. Temporal flicker is compared in our supplementary video (this is frame #47 of the MANIX semi-transparent sequence). Suffix NF and F mean using 'No Feature' and 'Feature' respectively. Ghosting and overblurring appear when 'No Feature' is used.

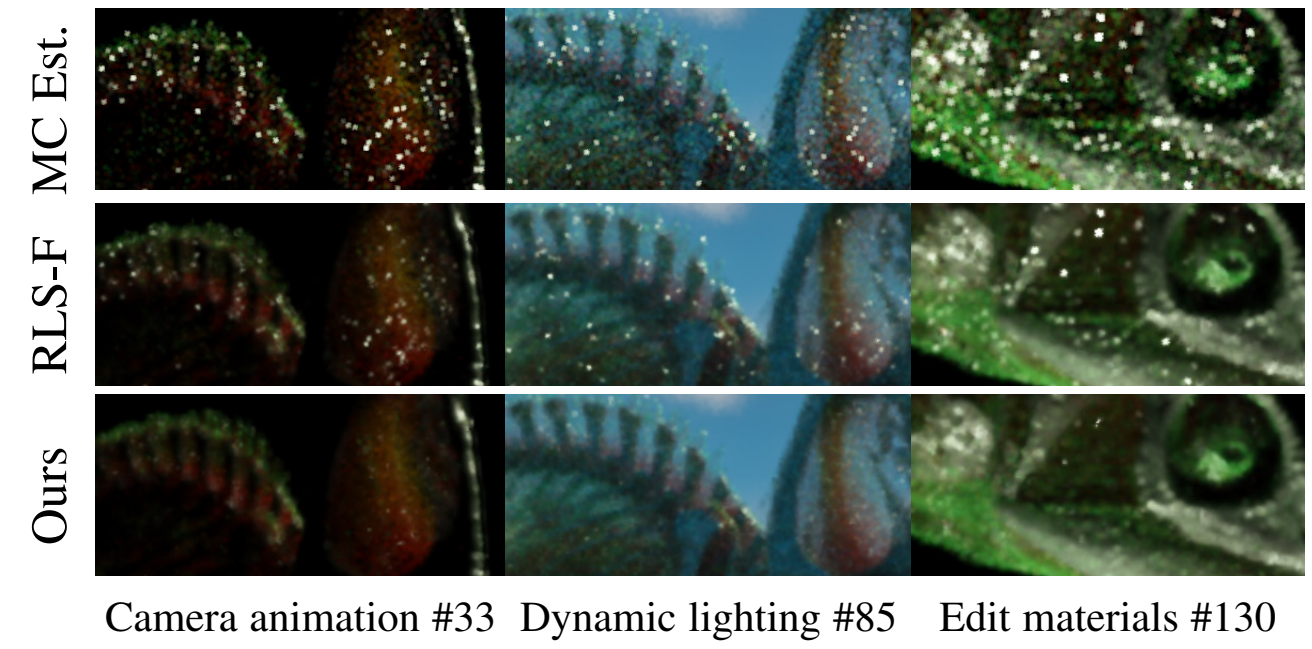

Figure 9: Results of RLS and wRLS both using our feature. While both methods produce reasonably good images thanks to our feature, wRLS suffers much less spike noise as it down-weights input pixel colors with high variances (check our supplemental video for temporal stability comparisons).

The equations to update the linear model coefficients and the inverse covariance matrix remains the same (see Eq. 9 and Eq. 11), with the exception that they use the new $\mathbf{q}_j^w(t)$ instead of $\mathbf{q}_j(t)$.

Fig. 8 shows the results of RLS and wRLS with and without our temporal feature. Both methods have reduced errors (see the bottom row in Fig. 8) when our temporal feature is used for the methods. While both techniques have much less noise than the MC estimated input, our method (wRLS) handles spike noise well compared to the RLS, as shown in Fig. 9, since the wRLS handles the heterogeneous noise adaptively by varying its weights.

**Spatio-temporal denoising using wRLS**. Our wRLS reduces the variance of VTP results by exploiting temporal coherence, but it is also desirable to use the spatial coherence between pixel colors within a frame to further reduce this variance. To this end, we apply a spatial filter to the denoised output of our temporal denoising. Technically, a linear model at a pixel can predict the colors of its neighboring pixels as well as its own pixel color using our per-pixel predictor vector. Specifically, to determine a final pixel color at $i$-th pixel, we blend colors predicted from its neighboring pixels defined by a $5 \times 5$ window centered at the $i$-th pixel. We have used a bilateral weight to average the multiple predictions and observed that this spatial filter reduces residual noise of the temporal output, without excessive blurring thanks to our feature-based linear predictions.

## 6 IMPLEMENTATION DETAILS

We have built our DVR denoising framework using the Exposure Render proposed in Kroes et al. [33]. Specifically, we have extended their framework to simulate multiple scattering effects and support additional light sources (e.g., point lights or HDR light probe captures). For example, we have utilized delta tracking for free-path sampling and built random walk paths to simulate multiple scattering effects. In particular, we have connected the last path vertex with a randomly selected light source using unidirectional sampling. We have used a separable Gaussian kernel to implement our pixel reconstruction filter, as this provides reasonable results for real-time purposes [50]. When computing our temporal feature, we have set the parameter $\alpha$ (in Eq. 12) to 0.75. Our DVR framework internally works in HDR (CIE-XYZ) colorspace, while for image display, it transforms final images into LDR (RGB) colorspace. In our framework, we perform the denoising in the original HDR space before applying any tone-mapping or gamma correction.

**Optical flow estimation**. An optical flow estimation is necessary to perform our temporal reprojection. In rendering, we typically exploit the world coordinates intersected by rays to estimate the per-pixel image velocity, commonly used in reprojection techniques for rendering surface models. Nevertheless, a ray can intersect at multiple locations within participating media due to its semitransparent nature. The naïve ways of handling such ambiguous world coordinates could select a world coordinate randomly or use an average one. We, however, have found that it could result in unstable temporal reprojections. To mitigate this problem, we compute our per-pixel image velocity using the closest world coordinate from the view position of the first real collision according to delta tracking. Our implementation of the optical flow estimation is a heuristic, but we have found that this simple choice works reasonably well for our tested scenarios (e.g., Fig. 8 and Fig. 14). We leave more principled manners of estimating optical flow in volume rendering as future work.

| Scene | Resolution (voxels) | Bps | Training data (only for RAE) |
|---|---|---|---|
| MANIX | $256 \times 256 \times 230$ | 16 | 500 frames |
| CHAMELEON | $1024 \times 1024 \times 1080$ | 8 | 500 frames |
| HELODERMA | $1024 \times 1024 \times 555$ | 8 | 500 frames |
| DRAGON | $1024 \times 1024 \times 1024$ | 8 | - |

TABLE 3: Scenes used in our experiments. We detail the volume resolution, the bits per sample (bps) and the number of images used to train DNN-based solutions like RAE [5]. Unless otherwise specified, all rendered images are always in HD resolution ($1280 \times$ 720p). The DRAGON dataset is only used at testing time.

## 7 RESULTS AND EVALUATION

In this section, we validate the performance of our DVR denoising using a series of experiments where we use a collection of stationary volume data sets: MANIX, CHAMELEON, HELODERMA and DRAGON (see Table 3). Our supplementary material contains a figure that shows the default transfer functions used in our experiments. Moreover, one of our supplementary videos demonstrates our denoising results during dynamic editing operations on

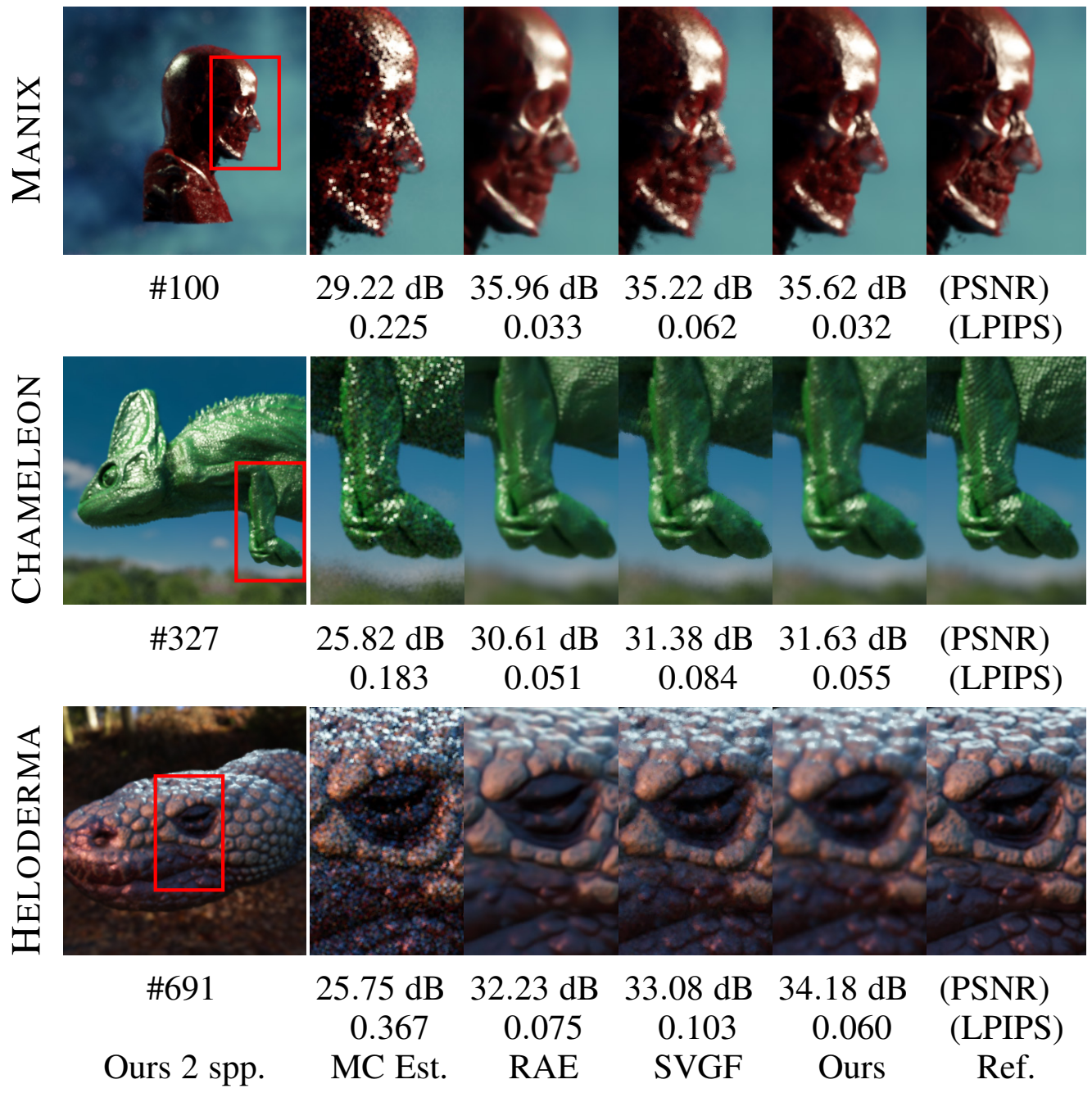

(a) Visual comparisons of denoising methods.

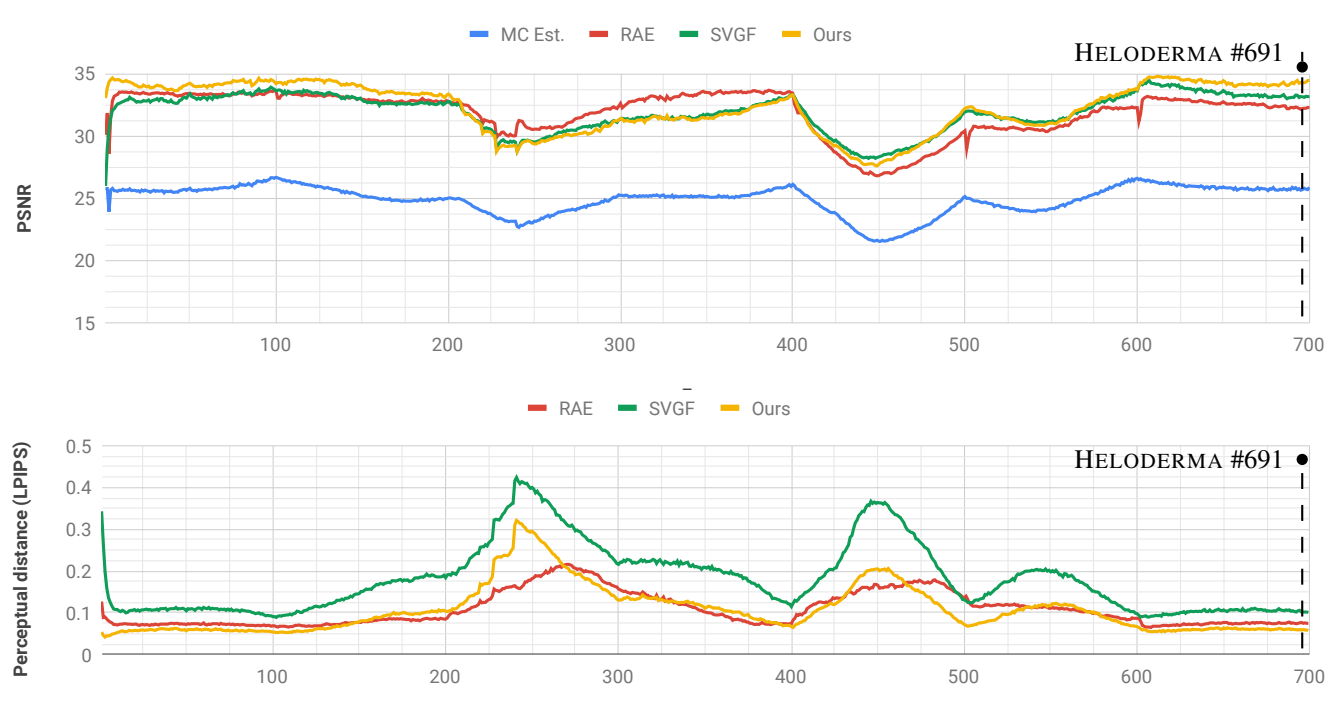

(b) Numerical accuracy over time for the HELODERMA sequence.

Figure 10: Real data sets tested for MC-DVR camera animations.

live recorded sessions running on a 2.8 GHz i7 Intel processor laptop with 32 GB of RAM and an Nvidia 1070 GTX GPU. All our testing scenes include dynamic changes, e.g., the camera viewpoint transformation, changing lights, or transfer functions. Unless stated otherwise, we used the same GPU configuration and screen resolution ($1280 \times 720$p) for all reported experiments.

### 7.1 Comparison with State-of-the-Art Denoisers

**Recurrent Auto-Encoders (RAE)**. We compared our method with a recent learning-based denoising proposed by Chaitanya et al. [5]. This previous work uses RAEs and was optimized for interactive reconstruction of MC image sequences with surface models, but it was not trained for volume rendering. We implemented their approach in Tensorflow [1] and retrained the network on DVR images for a fair comparison.

**Training RAEs for volume data.** We retrained their proposed RAE architecture using MC image sequences generated by our DVR. In volume rendering, auxiliary features can be very noisy [48], and thus our best option was to train RAEs using the input color buffer exclusively. For training the RAEs we generated 500 temporal sequences for MANIX, CHAMELEON, and HELODERMA scenes (as seen in Table 3). To generate the sequences for each scene, we modified the camera view and orientation around the volume, the light source position, and we also modified the transfer function of the volume. Note that we trained the network with the same scenes (MANIX, CHAMELEON, HELODERMA) used for our comparisons, just with some variations (e.g., camera view or the position of a light source). Furthermore, we trained a separate RAE network for each scene in order to maximize its denoising quality. The networks were trained for 150 epochs, taking one week of training time each using an Nvidia Quadro P6000 GPU. Note that we did not change the volume data itself for their testing (i.e., for comparisons with our method). While it is common practice to have different test scenes from training scenes, we chose this overly fair comparison so that our method can be compared with an upper bound performance of the RAE for the three scenes. Moreover, we used the DRAGON scene to compare our method with the RAE for general scenarios with a non-trained volume.

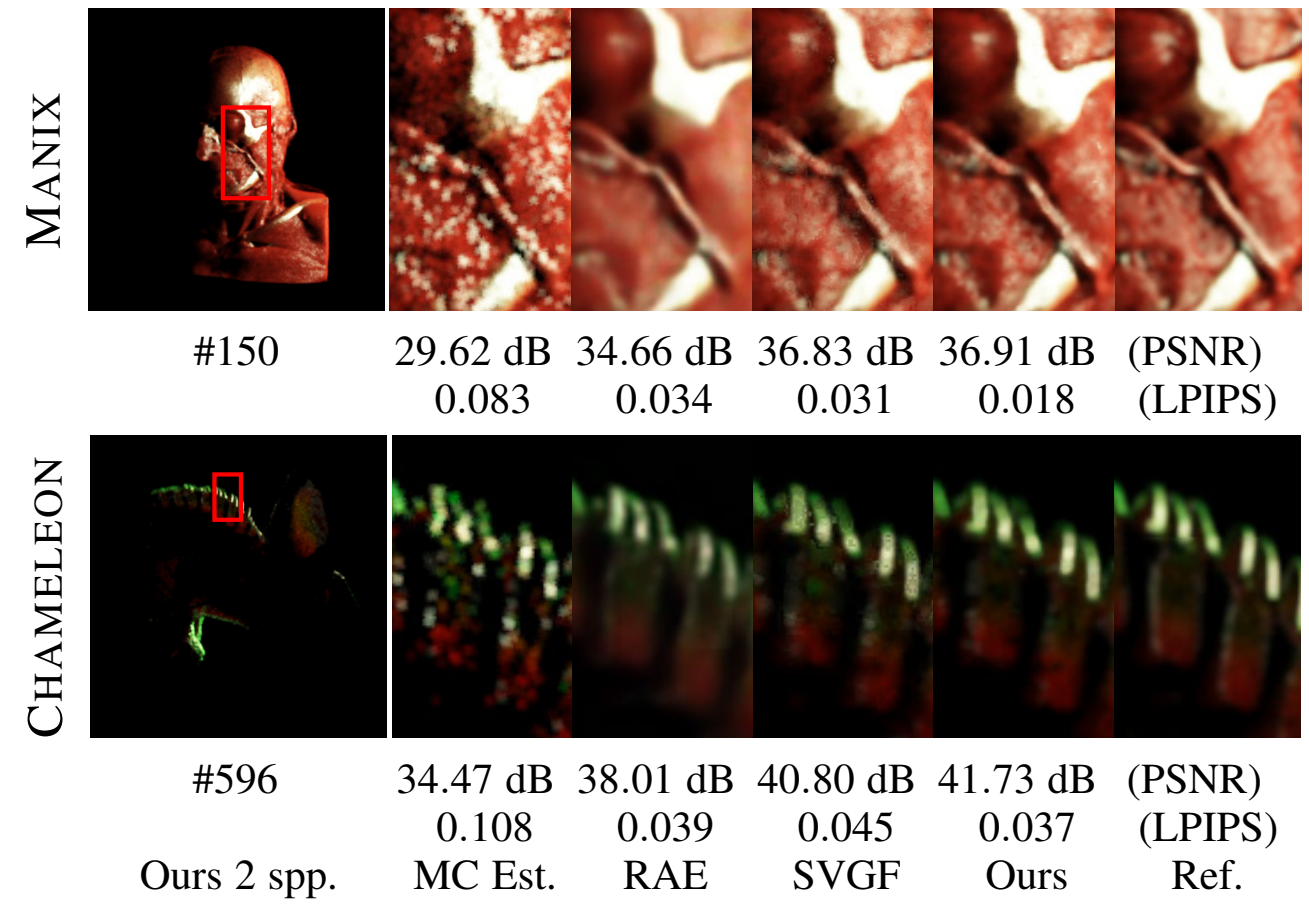

Figure 11: Results for MANIX and CHAMELEON tested under MC-DVR with animated point lights. Full time-wise comparisons, for both PSNR and LPIPS metrics, are available for all data sets in the supplemental material.

**Surface-based denoising methods**. We also compared with the spatiotemporal variance-guided filtering (SVGF) [53], whose primary purpose is to get rid of global illumination noise generated in scenes with surface models. We implemented SVGF following the guidelines in their paper, but we exchanged its reprojection scheme with ours (as explained in Sec. 6) in order to improve the denoising accuracy of SVGF for DVR.

**Evaluation metrics.** We report error images, computed as absolute per-pixel differences with a reference, peak-signal-to-noise ratio (PSNR) and learned-perceptual-image-patch-similarities (LPIPS) by Zhang et al. [60], which is a recent perceptual metric improving over SSIM and MS-SSIM. Usually a higher PSNR means higher quality, while for LPIPS a lower value means better quality.

### 7.2 DVR evaluation scenarios

We compared our method with the previous techniques given different types of user interaction scenarios for DVR. For each scenario, we computed a reference sequence with 1024 spp at 720p. Please, refer to our supplementary video and material for visual comparisons and time-wise PSNR and LPIPS metrics.

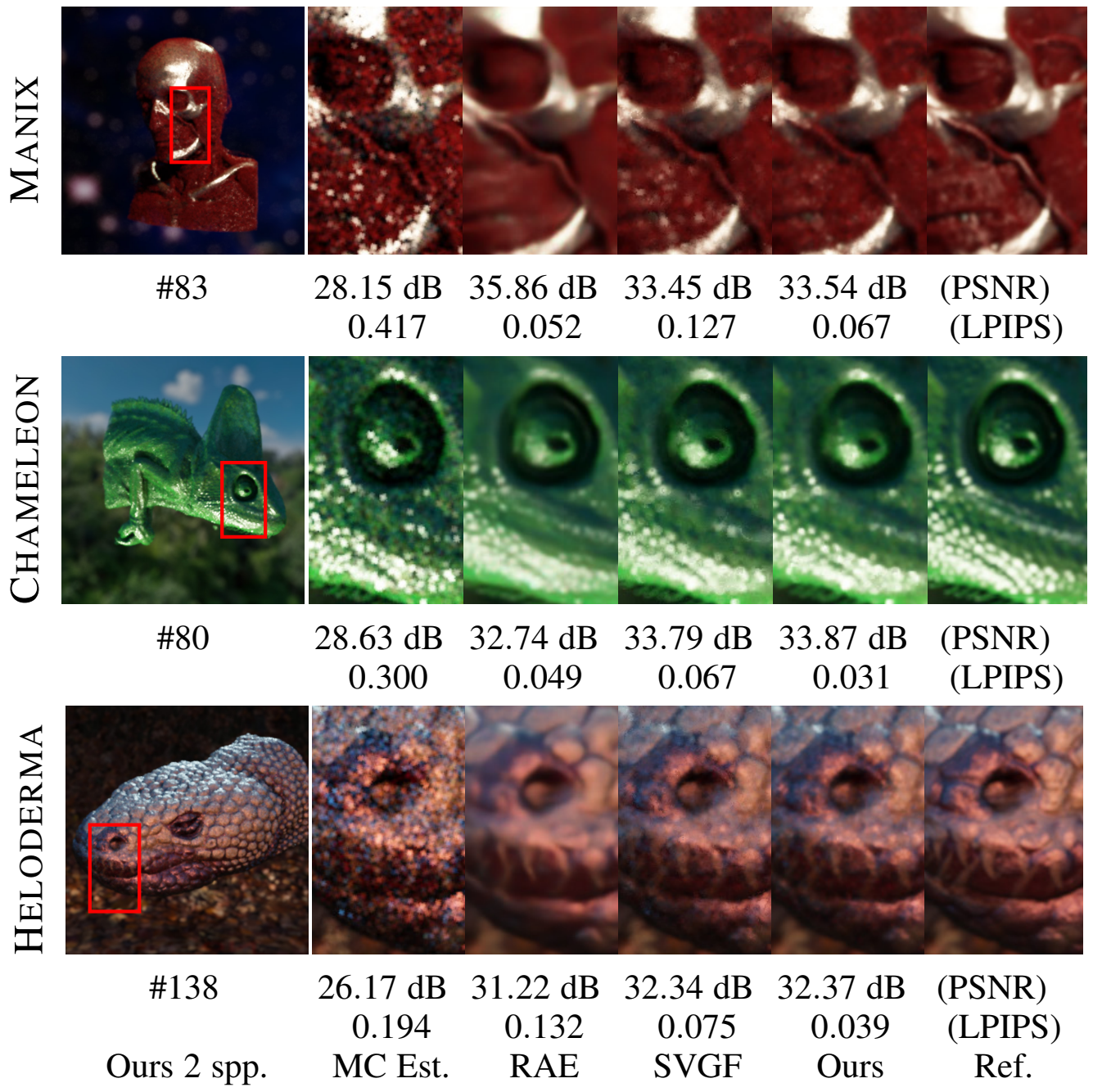

(a) Visual comparisons of denoising methods.

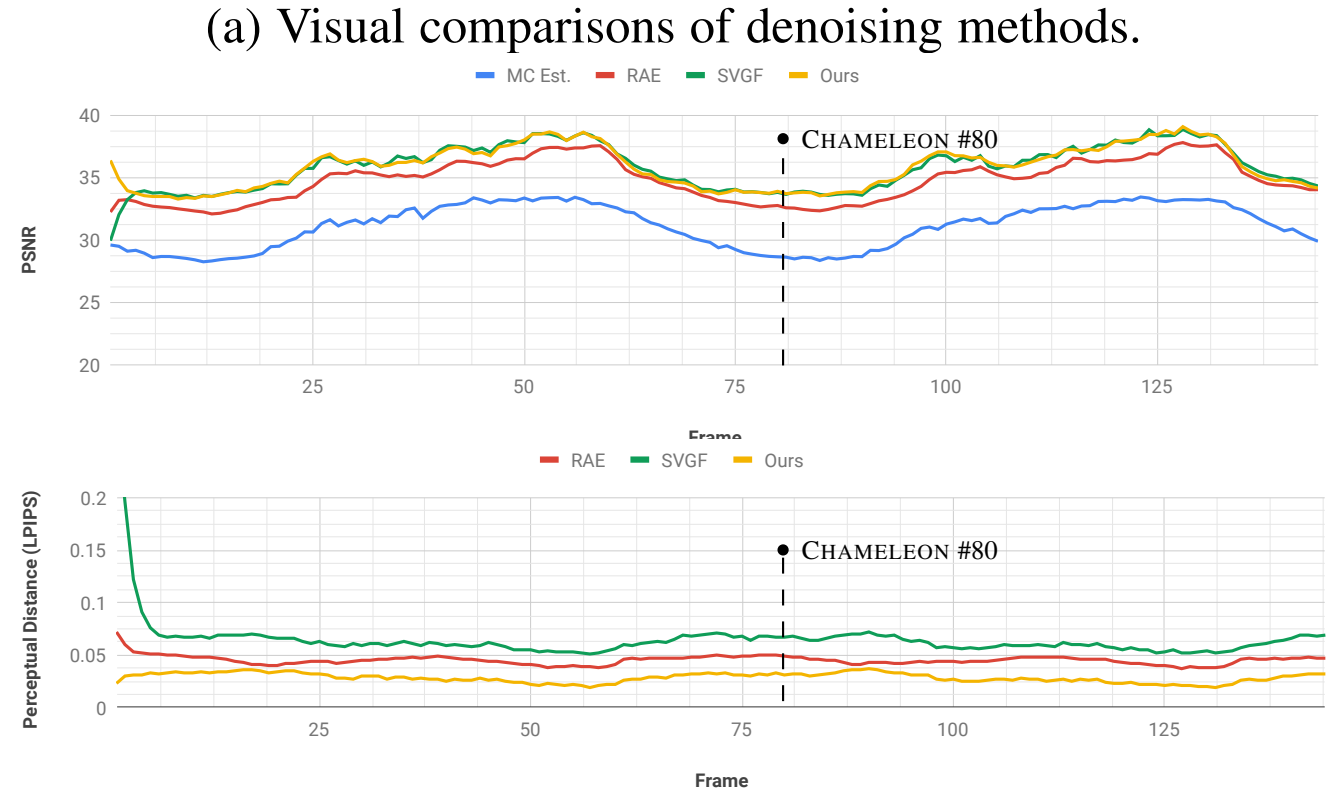

(b) Numerical accuracy over time for the CHAMELEON sequence.

Figure 12: Real volume data sets tested for MC-DVR with animated area lights.

**Scenario 1: modification of camera parameters**.
We created predefined camera animations that modified the position and orientation of the camera viewpoint for almost every frame during the sequence. Since our training data include simultaneous animations of various parameters, we decided to present a more straightforward case to the RAE so that the network can show a near-ideal denoising performance. We used the same camera animation presented at training time but fixing all other parameters, i.e., lighting and transfer function. We also run our method and SVGF on the same input sequences. Fig. 10 (a) shows visual comparisons of denoised frames during camera animations.

**Discussion.** In general, we noticed that our method exploits temporal coherence appropriately and produces temporally more stable results than RAE and SVGF. Our method behaves reasonably well either for surface reflections (e.g., skull reflections) or semi-transparent materials (e.g., MANIX eyelids or vessels). For still images, RAE and our method have a comparable visual quality (e.g., similar LPIPS in Fig. 10 (b)). While RAE can infer texture patterns quite convincingly (e.g., CHAMELEON skin bumps), its reconstruction is sometimes not as faithful and tends to wash out details. On the other hand, our method and SVGF tend to preserve the original lighting and geometric details better, but our method is temporally much more stable than SVGF.

**Scenario 2: interaction with light sources**.
We compared our method with RAE and SVGF under different lighting scenarios. It remains a highly impractical solution to reproduce all possible lighting conditions in the RAE training data. Thus, we decided to choose one representative lighting setup, like a rectangular area light as the principal light source and an environment map as a fill light. We generated training samples where the area light source is rotated around the main volume while we keep the environment lighting active. Analogously with the camera animation test, at testing time, we used the same path for the area light and the same environment lighting, but the camera and the transfer functions remained fixed. We tested our DVR denoising under different luminaire sources:

**Point lights.** Fig. 11 shows denoising results given a single point light animated around the main volume inside a dark environment. In this case, our method obtained comparable reconstruction quality as indicated by PSNR numbers but always better perceptual distance according to the LPIPS metric.

**Area lights.** For this test, we illuminated our data sets using the same HDR light probe captured environment used at training time and animated the rectangular area light rotating around the main volume. In Fig. 12 (a), we show visual comparisons with real volumes, and in Fig. 12 (b) the time-wise quantitative values (PSNR and LPIPS) for the CHAMELEON scene.

**Discussion.** In general, we observed that our technique produces visually sharper and more faithful reconstructions than RAE and SVGF approaches for this scenario. Regarding temporal coherence, our method provides more stable visualizations with less low-and-medium frequency flicker, as shown in our supplementary video. In the scenario with area lights, SVGF and our method generate sharper results than RAE. For example, high-frequency details on the thin vessels on the left temple of the head are well preserved by SVGF and ours. However, the SVGF suffers from spike noise, and its LPIPS numbers are much higher than with our method. This leads to the noticeable temporal flickering of SVGF, as shown in our supplementary video. While RAE shows better PSNR numbers than ours for MANIX, our method produces the best distortion and perceptual results for both CHAMELEON and HELODERMA. In particular, our LPIPS (0.039) is much lower than that of RAE (0.132), as RAE does not preserve the high-frequency details for HELODERMA. Also, our method produces temporally more stable results than the RAE for the tested scenes.

**Scenario 3: editing of transfer functions**.
Another interaction scenario is user manipulation of the transfer function, which is a classic interaction for visual inspection of volume data since it allows for hiding or enhancing different structures in the volume. The manipulation of the transfer function during interactive visualization would require recalculating the MC integral after every variation. To compare our method with RAE and SVGF, we used the same manipulation of the transfer functions as the one utilized to generate the training data for the RAE approach. The only variation for this test is that the camera

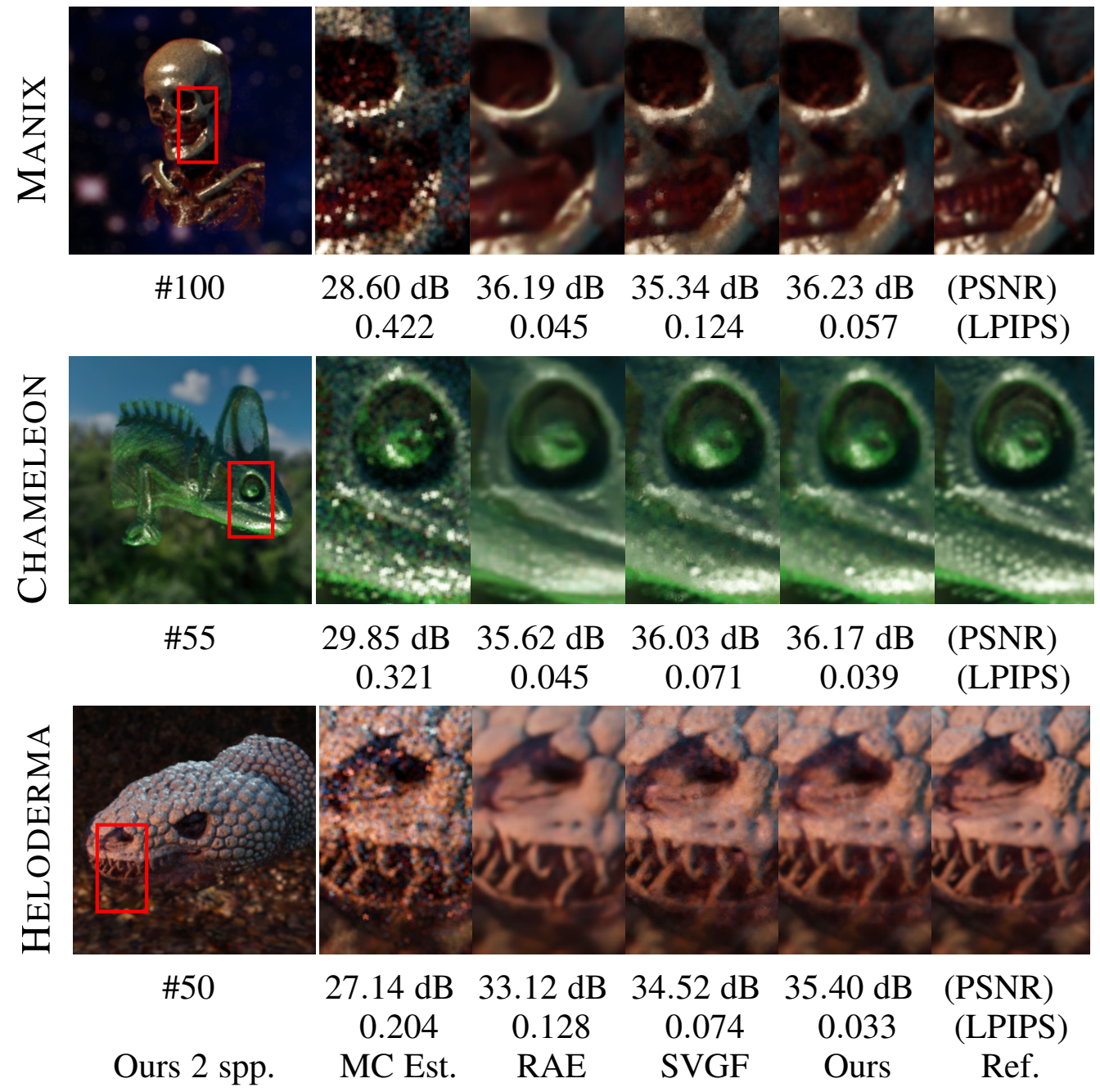

(a) Visual comparisons of denoising methods.

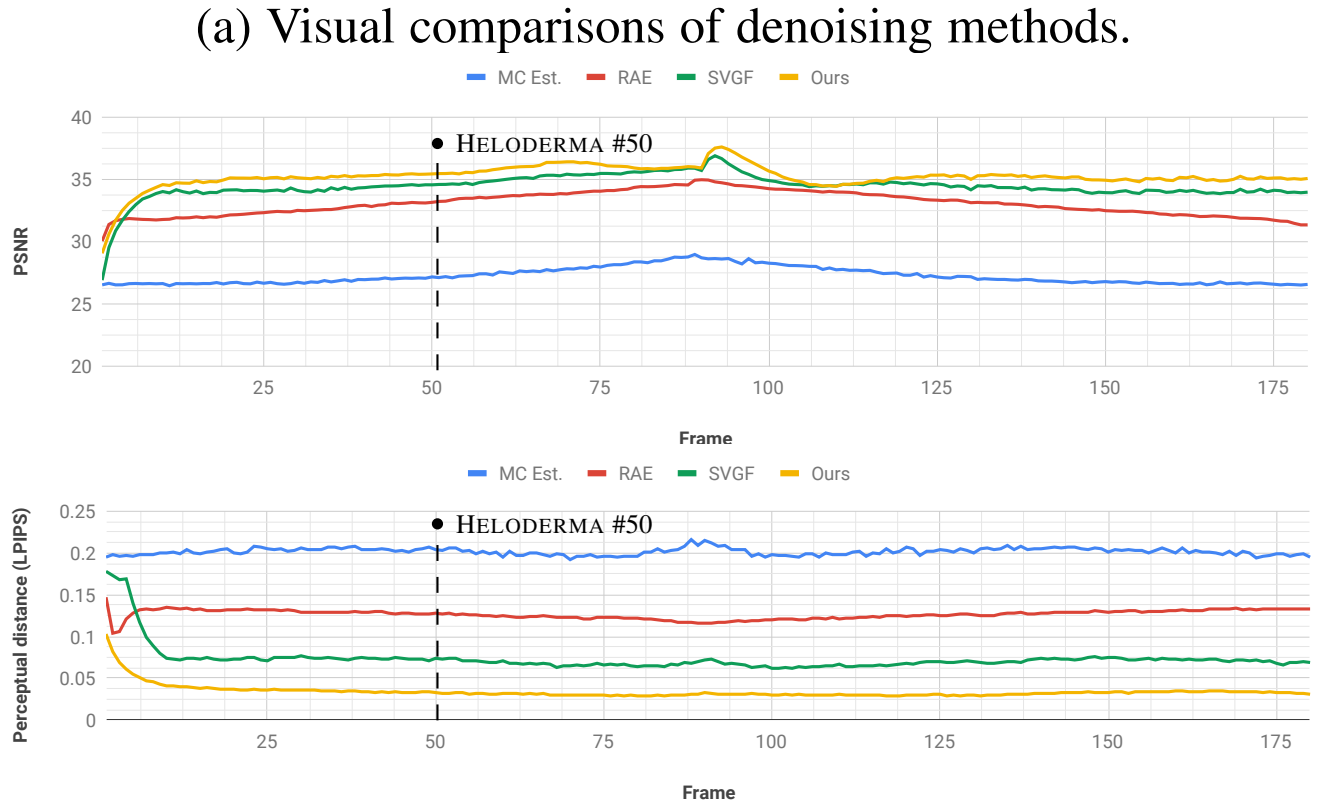

(b) Numerical accuracy over time for the HELODERMA sequence.

Figure 13: MC-DVR with transfer function variations in real-time.

and the lighting conditions are fixed.

**Discussion.** Fig. 13 (a) shows a single frame captured during the interaction to illustrate the visual quality obtained with each method, and Fig. 13 (b) shows time-wise PSNR and LPIPS comparisons for the HELODERMA volume. Our method often produces more accurate and perceptually preferable results than RAE and SVGF in terms of PSNR and LPIPS metrics. For example, the LPIPS of RAE (0.045) is 27% better than ours (0.057) for MANIX, but our technique outperforms RAE for the other cases. Overall, RAE tends to generate overly blurred results and the perceptual metric value of RAE (0.128) for HELODERMA is 3.9× higher than ours (0.033). As shown in our supplemental video, RAE generates low-and-medium frequency flickering artifacts and tends to blur high-frequency details of volumetric models. On the other hand, our method produces robust denoising results given different types of transfer functions.

### 7.3 DVR of highly-transparent volumes

We tested the denoising methods for complex semitransparent iso-surfaces and DVR camera animations (Fig. 14). Given these experiments, our technique shows comparable distortion and better perceptual errors than both RAE and SVGF, as shown in Fig. 14 (a) and the time-wise plots in Fig. 14 (b). In Fig. 15, we also tested semitransparent homogeneous media with the DRAGON data set, which was not used at training time by the RAE. When producing results for RAE, we always picked the best result from our three pre-trained networks.

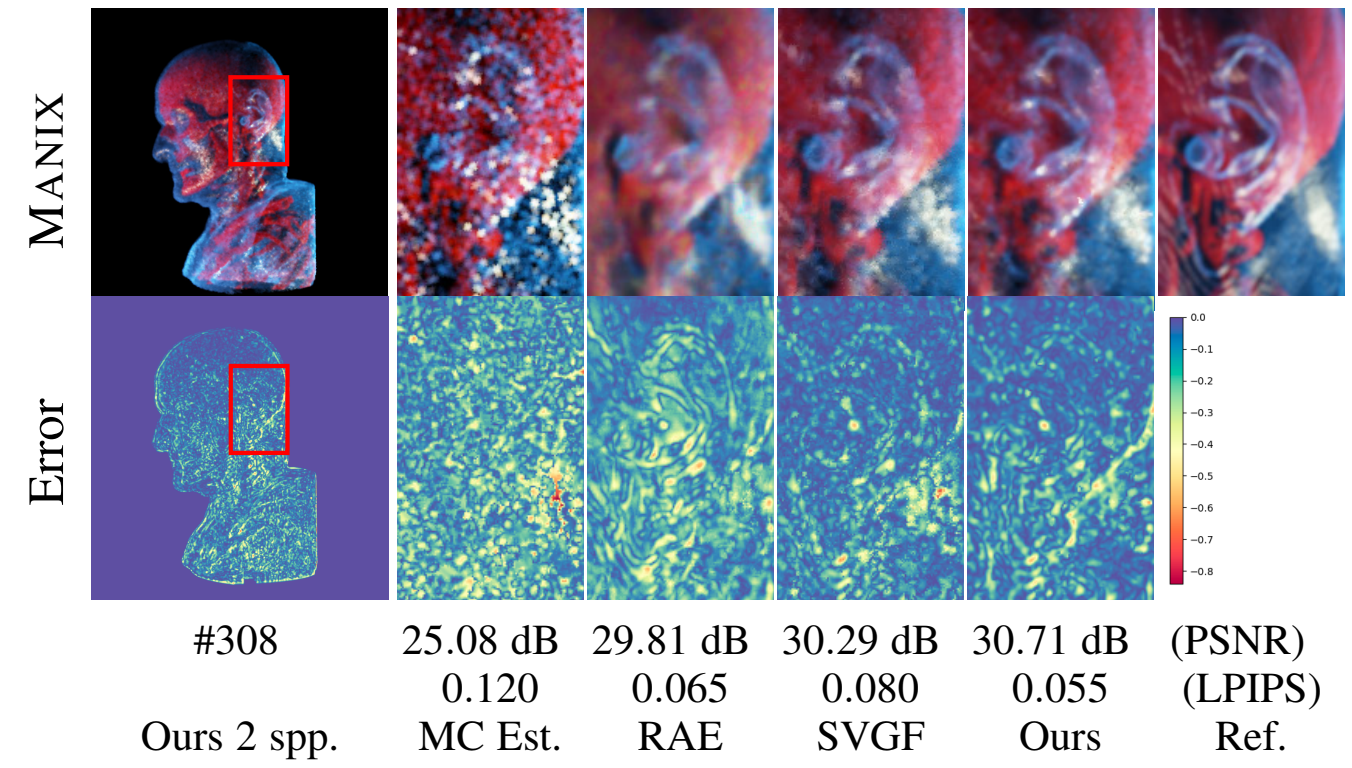

(a) Visual comparisons of denoising methods.

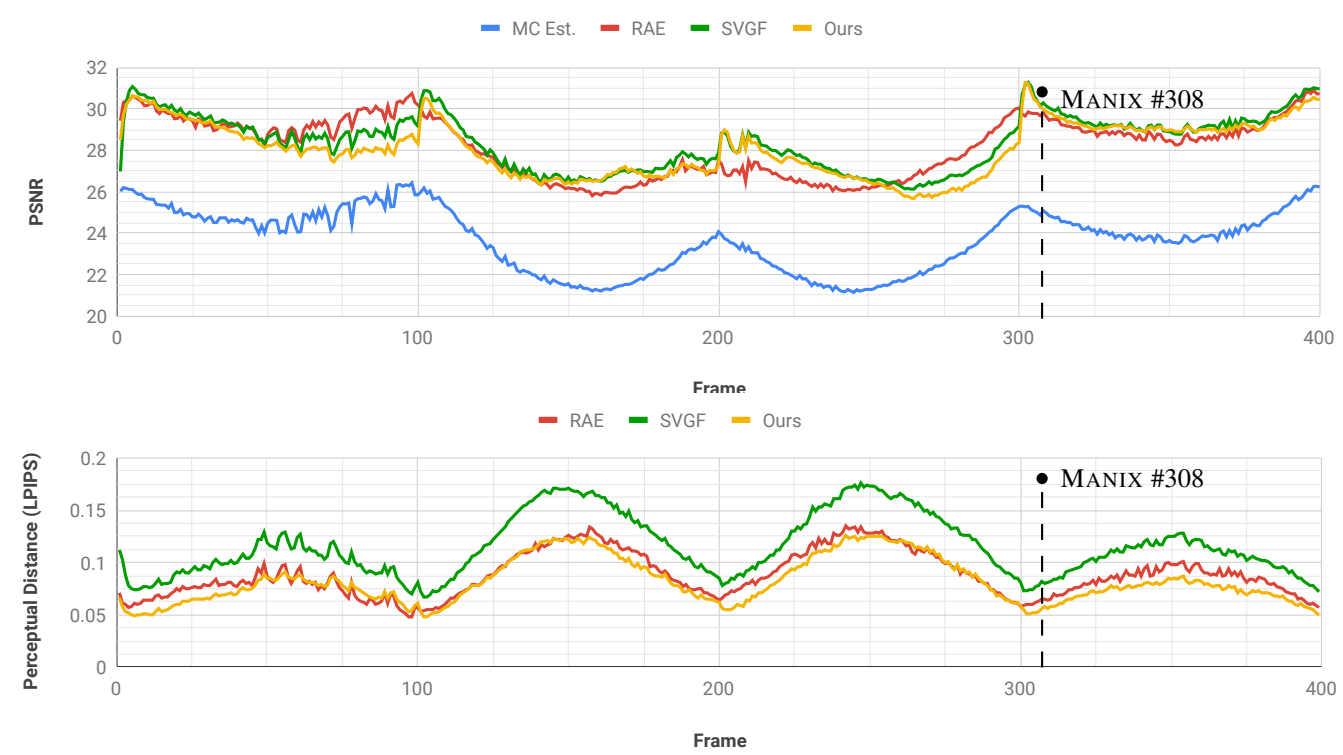

(b) Numerical accuracy over time for the semi-transparent MANIX .

Figure 14: Complex transfer functions demonstrating multiple semi-transparent isosurfaces on the MANIX data set. In general, our result achieves good reconstructions in comparison to RAE and SVGF, but much better temporal stability.

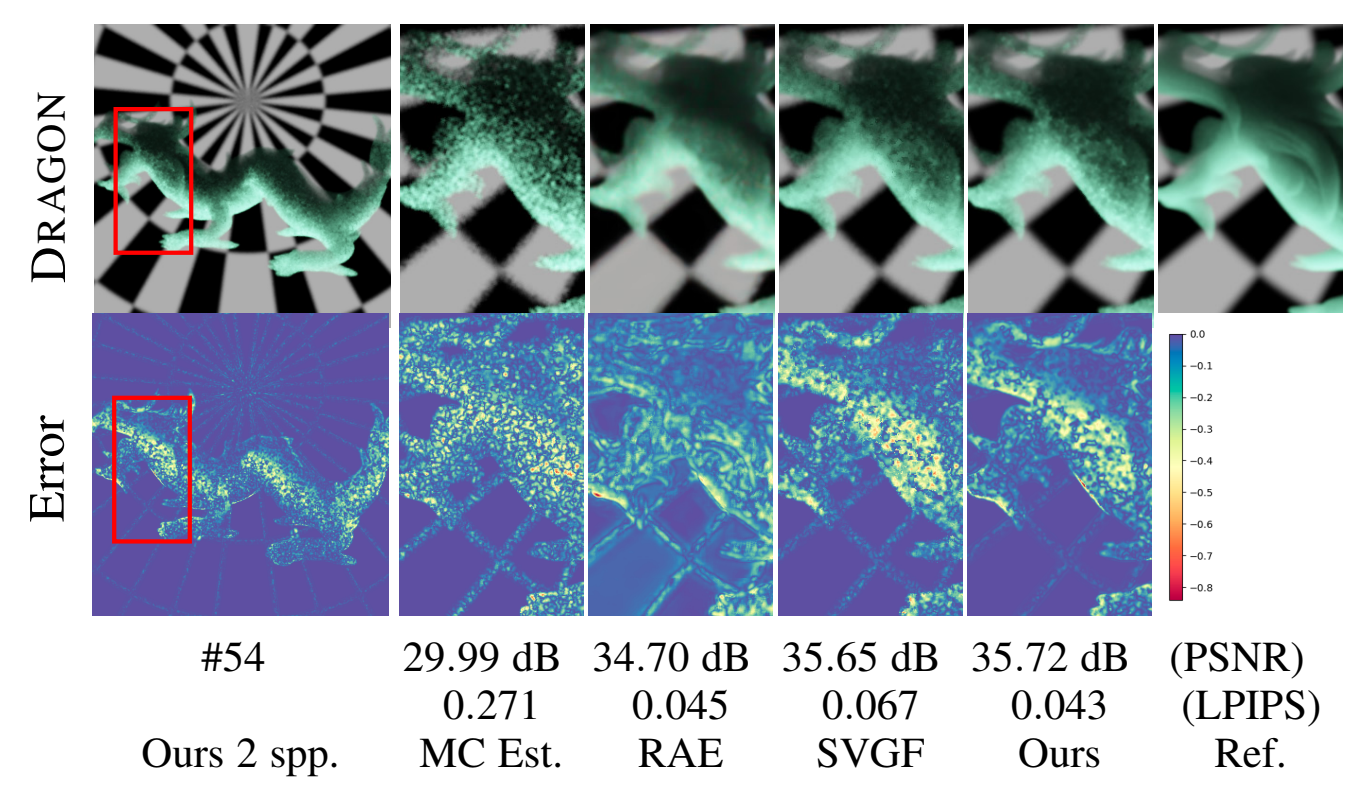

Figure 15: Complex geometry with dense homogeneous participating media material, shown on the DRAGON data set. Full time-wise comparisons, for both PSNR and LPIPS metrics, are available for this scene in the supplemental material.

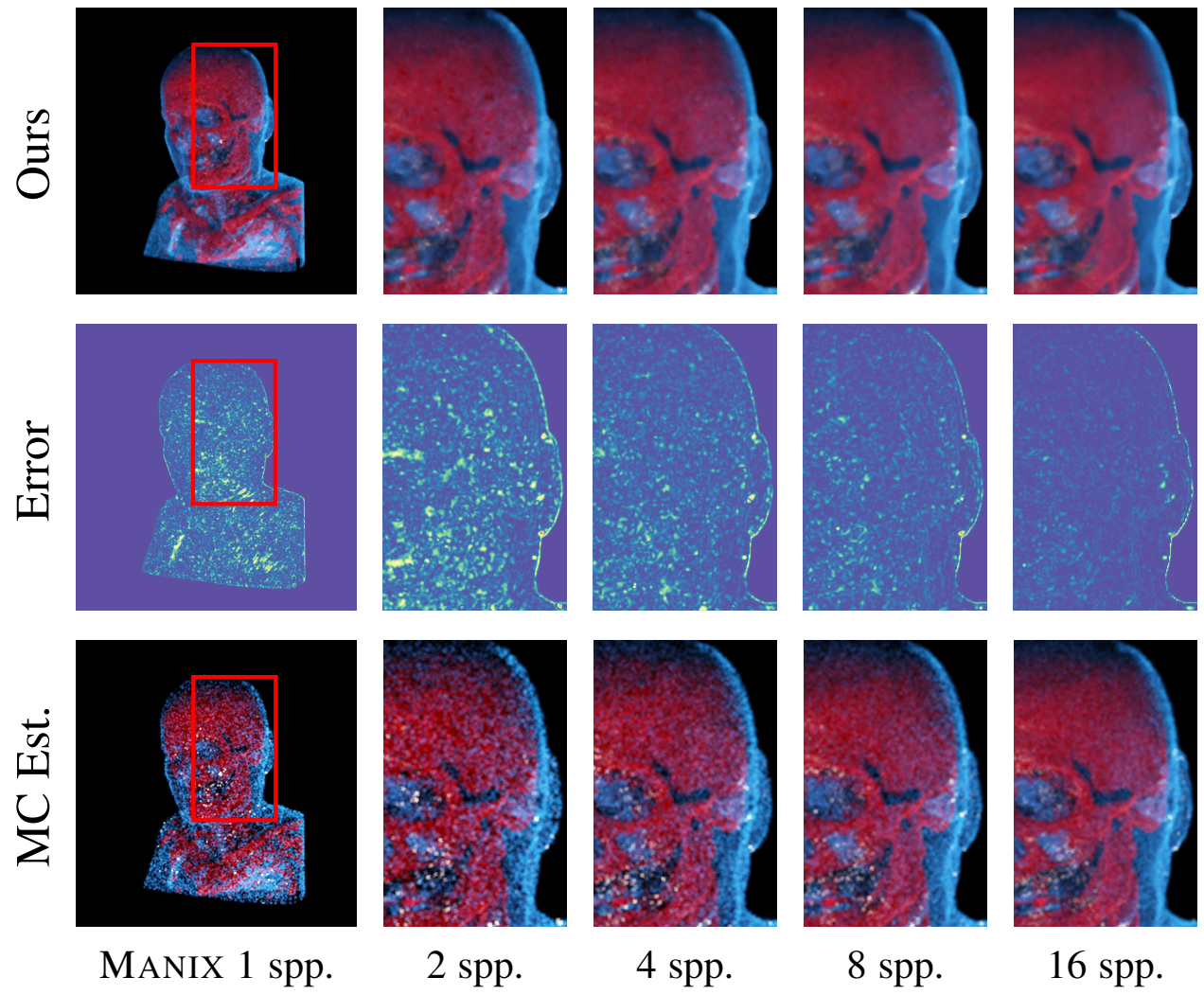

Figure 16: Our results with different numbers of samples. We render semi-transparent multiple isosurfaces with narrow transfer function bands for the MANIX dataset.

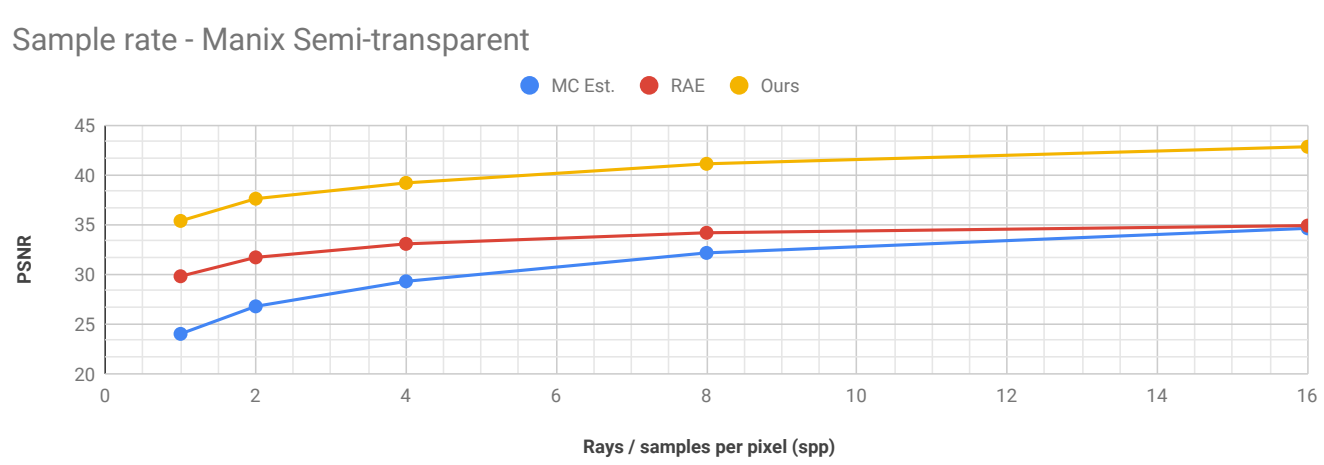

Figure 17: Numerical convergence of RAE and our method with different numbers of samples for the semi-transparent MANIX dataset.

### 7.4 Convergence and temporal stability

**Convergence.** Fig. 16 shows our results with varying the samples per pixel (spp), and it indicates that our numerical accuracy improves progressively as increasing the spp. We also compare our numerical convergence with RAE in Fig. 17, and our method shows better convergence than the previous method. For example, the improvements of RAE over noisy input images become more modest, as the sample count increases. On the other hand, our method consistently improves the input images.

**Temporal stability.** The temporal stability of animated sequences is a critical aspect of maximizing user experience during interactive visualization. As demonstrated in our supplemental videos, our approach provides more consistent temporal stability while minimizing the negative impact of outlier samples (i.e., spike noise) thanks to our wRLS. Camera animations are challenging scenarios for our temporal reprojection, but we did not notice strong overblur or ghosting in our experiments, especially when compared to RAE that suffers from low-and-medium frequency flickering artifacts. In Fig. 18, we tested the denoising methods for a static camera but in the presence of input temporal MC variance caused by only the random jittering present in the camera rays. Our method handles spike noise appropriately while preserving the details of the volume.

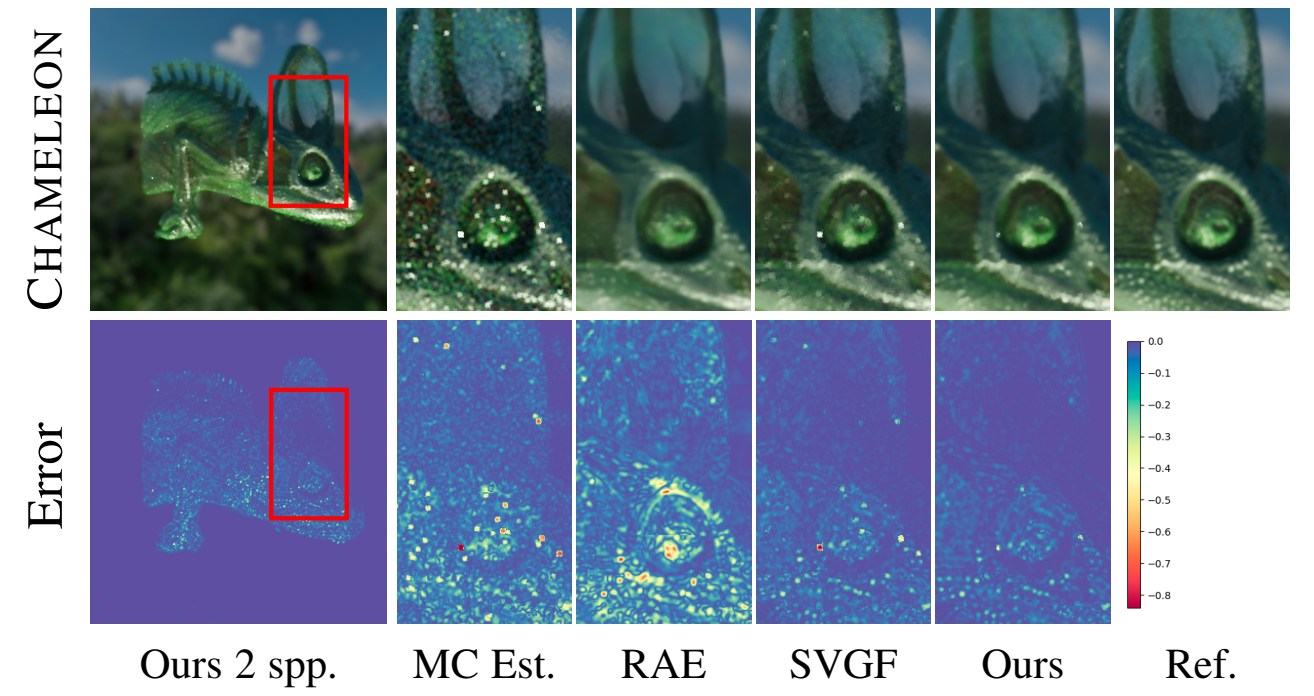

Figure 18: Temporal flicker and outlier removal on the CHAMELEON data set. Here we use a highly transparent transfer function and one extra scattering bounce per sample. This complex setup creates more outlier samples and stresses the denoisers.

### 7.5 Runtime performance

The computational complexity of our denoising mainly depends on image resolution (e.g., $1280 \times 720$p HD used for our tests). In our experiments, the execution time allocated to MC sampling and DVR rendering with 2 spp is $\approx 40$ ms. The average runtime of our denoiser is more than $\times 3$ less, being around $\approx 12$ ms per frame ($\pm 2.5$ ms). The computation overhead of our denoising remains solely dependent on the image resolution, and thus 12 ms can be well suited for real-time DVR. Also, our method and SVGF do not require expensive preprocessing stages, as it happens for RAEs where network training can take several days. The running times for RAE and SVGF are much higher than the overheads described in the original papers. Our RAE implementation works offline and takes over a second to process a single frame. Our SVGF implementation working with DVR can take $\approx 150$ ms per frame, which is also much higher than the overhead described in their paper. Our unoptimized implementations may cause all this, and thus, for reference purposes, we report the optimized times in their papers. We took as reference timings the performance reported in authors' original papers. RAE and SVGF required 54.9 ms and $\approx$ 4.4 ms respectively for the same 720p HD resolution.

### 7.6 Limitations and future work

The proposed method makes use of temporal reprojection to identify candidate linear models for the next predictions. However, this reprojection can make an error, especially inside heterogeneous volumes. In our experiments, we noticed incorrect reprojections could result in a small degree of overblur or sporadic ghosting when strong disocclusions occur, and in this case, the linear models can fail to compensate for this change. This trade-off falls within the expectations of any temporal filter since small amounts of overblurring are often perceived as more acceptable than the flickering effect [58], especially when flickering is significantly reduced by the filtering method. Nevertheless, it would be ideal for designing robust reprojection schemes specialized for heterogeneous volumes, and we leave that as future research. Our method approximates radiance changes over time with linear functions, and the approximation error can become large when the radiance varies in a totally non-linear manner. Fig. 19 shows a challenging scenario where a dynamic light changes its color over time. Our method produces improved results compared to RAE and SVGF. However, to further optimize our denoising, we would like to investigate an automatic parameter selection for the forgetting factor. It would also be interesting to integrate our denoising into an advanced framework that handles volumes with non-exponential free-flight distributions [3].

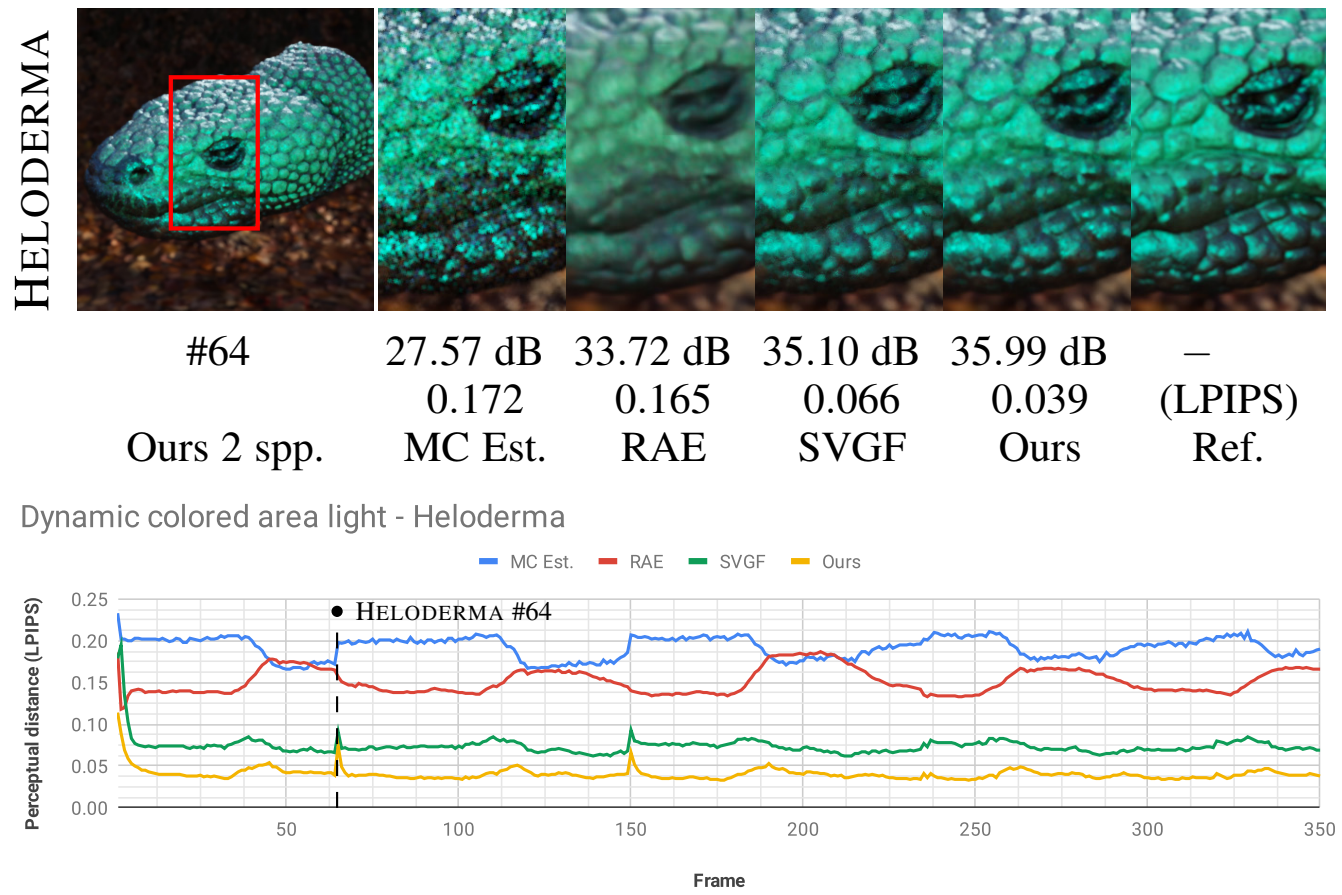

Figure 19: A challenging scenario with an animated area light that changes its color rapidly over time. RAE shows a color shift and excessive blur on this difficult scenario. Our method also generates slightly over-blurred results but shows better numerical reconstruction and more pleasant visual results than state-of-the-art.

## 8 Conclusions

This paper has presented a novel real-time denoising technique that reduces noise in VPT when used for DVR. In particular, our method produces high visual fidelity and temporally stable results for challenging scenarios where VPT suffers severe noise due to the real-time constraints (e.g., 1 or 2 spp). Technically, we reduce the variance of VPT effectively using per-pixel linear model predictions and additional spatial filtering, which take advantage of the spatio-temporal coherence among pixel colors in consecutive frames. Our denoising achieves temporal stability thanks to our weighted recursive least squares that addresses heterogeneous noise introduced by VPT. We have extensively demonstrated that our framework enables users to manipulate volume data experiencing much less temporal flicker interactively.

## Acknowledgments

The authors would like to thank reviewers for their insightful feedback. We also thank S. Oh, F. Lumbreras and J. Serrat for their help with Tensorflow, and T. Kroes for Exposure Render. The datasets are courtesy of the OsiriX Foundation (MANIX), University of Texas (CHAMELEON), University of Arizona (HELODERMA) and XYZ RGB Inc. (DRAGON). This work has received funding from the European Union's Horizon 2020 research and innovation programme under the Marie Skłodowska-Curie g.a. No 665919. J.A. Iglesias-Guitian acknowledges the UDC-Inditex InTalent programme, the Spanish Ministry project TIN2017-88709-R, RYC2018-025385-I (MCIU/AEI/FEDER, EU), FEDER Galicia ED431G 2019/01 and the Nvidia GPU Grant Program. B. Moon was supported by The Cross-Ministry Giga KOREA Project grant funded by the Korea government (MSIT) (No. GK20P0300).

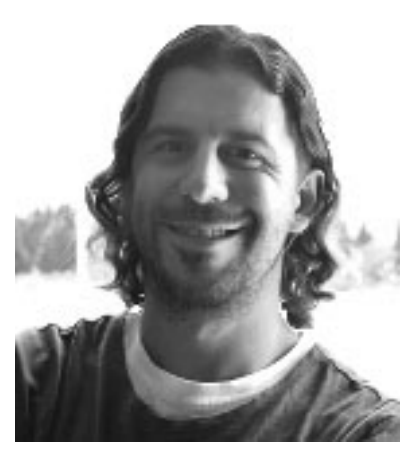

**Jose A. Iglesias-Guitian** is a Ramon y Cajal researcher funded by the InTalent Programme at UDC (University of A Coruña) and CITIC (Centre for ICT Research). Before, he was with the Computer Vision Center and the Universitat Autònoma de Barcelona (UAB) as a Marie Curie fellow. Before coming back to Spain he was with Disney Research. He received his Ph.D. degree in Electronics and Computer Engineering in Italy (2011). He is EG and ACM member.

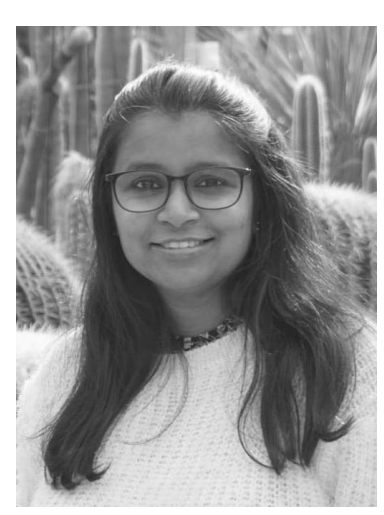

**Prajita Mane** has completed a M.S. at GIST (Gwangju Institute of Science and Technology). She received her B.E. degree in Electronics and Telecommunications from Pune University in 2010. Her research interests include visualization, interactive rendering and deep learning.

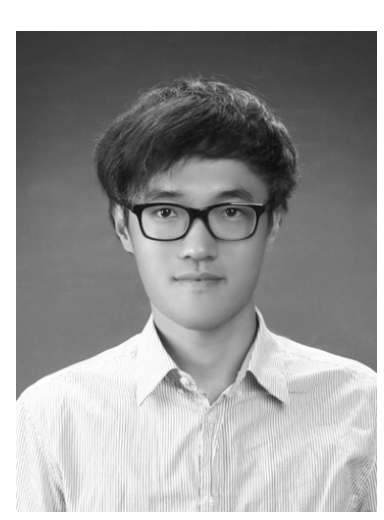

**Bochang Moon** is an assistant professor at GIST (Gwangju Institute of Science and Technology). He received his M.S. and Ph.D. degrees in computer science from KAIST in 2010 and 2014, respectively. He was a postdoctoral researcher at Disney Research. His research interests include rendering, denoising, and augmented and virtual reality. He served as a PC member for international conferences such as EGSR, I3D, PG and CGI. He is a member of IEEE and ACM.